\documentclass[11pt,a4paper]{article}
\usepackage[colorlinks]{hyperref}
\usepackage[utf8]{inputenc}
\usepackage[english]{babel}
\usepackage{amsmath}
\usepackage{amsfonts}
\usepackage{amssymb}
\usepackage{graphicx}
\usepackage{lmodern}
\usepackage{tabulary}
\usepackage{multirow}
\usepackage{color}
\usepackage[margin=2.5cm]{geometry}
\usepackage[affil-it]{authblk}

\title{Linear models for the impact of order flow on prices \\ I.  Propagators: Transient vs. History Dependent Impact}
\author[1]{Damian Eduardo Taranto}
\author[2]{Giacomo Bormetti}
\author[3]{Jean-Philippe Bouchaud}
\author[1]{\\Fabrizio Lillo}
\author[3]{Bence T\'oth}

\affil[1]{Scuola Normale Superiore, Piazza dei Cavalieri 7, 56126 Pisa, Italy}
\affil[2]{Department of Mathematics, University of Bologna, Piazza di Porta San Donato 5, 40126 Bologna, Italy}
\affil[3]{Capital Fund Management, 23-25, Rue de l'Universit\'e 75007 Paris, France}
\vspace{0.5cm}
\begin{document}
\maketitle
\begin{abstract}
  Market impact is a key concept in the study of financial markets and several models have been proposed in the literature so far. The Transient Impact Model (TIM) posits that the price at high frequency time scales is a linear combination of the signs of the past executed market orders, weighted by a so-called propagator function. An alternative description -- the History Dependent Impact Model (HDIM) -- assumes that the deviation between the realised order sign and its expected level impacts the price linearly and permanently. The two models, however, should be extended since prices are a priori influenced not only by the past order flow, but also by the past realisation of returns themselves. In this paper, we propose a two-event framework, where price-changing and non price-changing events are considered separately. Two-event propagator models provide a remarkable improvement of the description of the market impact, especially for large tick stocks, where the events of price changes are very rare and very informative. Specifically the extended approach captures the excess anti-correlation between past returns and subsequent order flow which is missing in one-event models. Our results document the superior performances of the HDIMs even though only in minor relative terms compared to TIMs. This is somewhat surprising, because HDIMs are well grounded theoretically, while TIMs are, strictly speaking, inconsistent.
\end{abstract}

\section{Introduction}
Understanding how the order flow affects the dynamics of prices in financial markets is of utmost importance, both from a theoretical point of view (why and how prices move?) and for practical/regulatory applications (i.e trading costs, market stability, high frequency trading, `Tobin' taxes, etc.). The availability of massive data sets has triggered a spree of activity around these questions~\cite{hasbrouck88, hasbrouck91,jones,biais,engle,cont,bacry14} (for a review see \cite{bouchaudetal2009}). One salient (and initially unexpected) stylized fact is the long-memory of the order flow, i.e. the fact that buy/sell orders are extremely persistent, leading to a slowly decaying correlation of the sign of the order imbalance~\cite{bouchaud2004,lillo2004}. This immediately leads to two interesting questions: first, why is this so? Is it the result of large ``metaorders'' being split in small pieces and executed incrementally, or is it due to herding or copy-cat trades, i.e. trades induced by the same external signal or by some traders following suit, hoping that the initial trade was informed about future price movements? Second, how is it possible that a highly predictive order flow impacts the price in such a way that very little predictability is left in the time series of price changes? 

Several empirical investigations, as well as order of magnitude comparisons between the typical total size of metaorders and the immediately available liquidity present in the order book, strongly support the ``splitting'' hypothesis \cite{lillo2005,toth2015}. Since the metaorder has to be executed over some predefined time scale (typically several days for stocks), the structure of the order flow is expected to be, in a first approximation, independent of the short term dynamics of the price and can be treated as exogenous -- see below. The idea then naturally leads to a class of so-called ``propagator'' models, where the mid-point price $m_t$ (just before trade at time $t$) can be written as a linear superposition of the impact of all past trades, considered as given, plus noise \cite{bouchaud2004,bouchaud2006}:
\begin{equation}
  m_t= \sum_{t'<t} \left[ G(t-t') \epsilon_{t'} + \eta_{t'} \right] + m_{-\infty}
  \label{eqn:prom_price_process}
\end{equation}
where $\epsilon_{t'}$ is the sign of trade at time $t'$ ($\pm 1$ for buy/sell market orders), $\eta_{t'}$ is a noise term which models any price changes not induced by the trades (e.g. limit orders/cancellations inside the spread, jumps due to news, etc.). The function $G(\ell)$ is called the ``propagator'' and describes the decay of impact with time. The crucial insight of this formulation is precisely that this impact decay may counteract the positive auto-correlation of the trade signs and eventually lead to a diffusive price dynamics (see \cite{bouchaud2004} and below). Although highly simplified, the above framework leads to an interesting approximate description of the price dynamics. Still, many features are clearly missing, see \cite{eisler2012a}:
\begin{itemize}
\item First, the above formalism posits that all market orders have the same impact, in other words $G$ only depends on $t-t'$ and not on $t$ and $t'$ separately, which is certainly very crude. For example, some market orders are large enough to induce an immediate price change, and are expected to impact the price more than smaller market orders. One furthermore expects that depending on the specific instant of time and the previous history, the impact of market orders is different. 
\item Second, limit orders and cancellations should also impact prices, but their effect is only taken into account through the time evolution of $G(\ell)$ itself that phenomenologically describes how the flow of limit orders opposes that of market orders and reverts the impact of past trades. 
\item Third, the model assumes a {\it linear} addition of the impact of past trades and neglect any non-linear effects which are known to exist. For example, the total impact of a metaorder of size $Q$ is now well known to grow as $\approx \sqrt{Q}$, a surprising effect that can be traced to non-linearities induced by the deformation of the underlying supply and demand curve, see e.g. \cite{toth2011,mastromatteo2014,donier2015}.
\end{itemize}
However, before abandoning the realm of linear models, it is interesting to see how far one can go within the (possibly extended) framework of propagator models, in order to address point 1 and 2 above. 

The aim of this work is to explore generalised linear propagator models, in the spirit of \cite{eisler2012a}, with a fully consistent description of the impact of different market events and of the statistics of the order flow. For the sake of readability, we have decided to present our results in two companion papers. In the present first part (I), we investigate in detail two possible generalisations of Eq. \ref{eqn:prom_price_process} above, where price-changing and  non price-changing market orders are treated differently. We show that separating these two types of events already leads to a significant improvement of the predictions of the model,  in particular for large tick stocks. We revisit the difference between the ``transient impact model'' (TIM) and the ``history dependent impact model'' (HDIM) introduced in \cite{eisler2012b}, correct some misprints in that paper, and show that HDIM is always (slightly) better than TIM for small tick stocks, as expected intuitively. We then turn to the modelling of the order flow in the companion paper (II), with in mind the necessity of keeping the linearity of the predictors of future order flow, as assumed in HDIMs. The so-called Mixed Transition Distribution (MTD) model is a natural framework for constructing a versatile time series model of events, with a broad variety of correlation structures~\cite{raftery1985,berchtold1995}. %\textcolor{red}{BENCE: I removed the detailed discussion of paper II.}
%In the case of two event types (such as $\pm 1$ strings of market orders, or $\mathrm{NC}, \mathrm{C}$ strings of events), the MTD reduces to the ``Discrete Autoregressive" (DAR) model. To our knowledge, our work is the first attempt to calibrate an MTD model to financial time series (excluding \cite{taranto2014} where a DAR model is calibrated to $\pm 1$ market orders). The second part of our work is therefore mostly methodological; we discuss maximum likelihood or generalized method of moments as ways to calibrate these high-dimensional models. We show that while MTD fares quite well at accounting for the time series of the events $(\epsilon_t,\pi_t)\in\lbrace(-1,\mathrm{C})\rbrace$, $\lbrace(-1,\mathrm{NC})\rbrace$, $\lbrace(+1,\mathrm{NC})\rbrace$, $\lbrace(+1,\mathrm{C})\rbrace$ (see below for the formal definition) in the case of small tick stocks, some irreducible discrepancies persist for large tick stocks. We attempt to explain why MTD models are insufficient in this case.

%\section{The simplest propagator model: A short review }\label{sec:propagator}
\section{The one-event propagator model}
\label{sec:propagator}
The propagator model defined by Eq. \ref{eqn:prom_price_process} above can alternatively be written in its differential form, where instead of the price process we consider the return process,
$r_t = m_{t+1} - m_t$:
\begin{equation}
  r_t=G(1)\epsilon_t+\sum_{t'<t}{\cal G}(t-t') \epsilon_{t'}+\eta_t , \qquad  {\cal G}(\ell)\equiv G(\ell+1)-G(\ell),
  \label{eqn:prom_return_process}
\end{equation}
where $G(\ell \leq 0) \equiv 0$. In the following we will call this model Transient Impact Model (as in \cite{eisler2012b}) and we label the predicted values according to the above model with TIM1 where the "1" refers to the fact that one propagator function, $G(\ell)$, characterizes the model.

Empirical results show \cite{bouchaud2004, eisler2012a} that for small ticks $G(\ell)$ is a decreasing function with time, therefore the kernel ${\cal G}(\ell>0)$ is expected to be a negative function. This means that the impact of a market order is smaller if it follows a sequence of trades of the same sign than if it follows trades of the opposite sign. The authors of \cite{lillo2004} call this behaviour the ``asymmetric liquidity'' mechanism:  the price impact of a type of order (buy or sell) is inversely related to the probability of its occurrence. The reason for this mechanism is that liquidity providers tend to pile up their limit orders in opposition of a specific trend of market orders \cite{bouchaud2006,mastromatteo2014}, whereas liquidity takers tend to reduce the impact of their trades by adapting their request of liquidity to the available volume during the execution of their metaorders \cite{taranto2014}.
%\textcolor{red}{BENCE: TOO DENSE!} \textit{The dynamic nature of the liquidity providing and the liquidity taking in financial markets is responsible for this mechanism: Liquidity providers tend to pile up their limit orders in opposition of a specific trend of market orders, the so called ``stimulated liquidity refill'' \cite{bouchaud2006,mastromatteo2014}, whereas the liquidity takers tend to reduce the impact of their trades by reducing the request of liquidity during the execution of their metaorders \cite{taranto2014}. }

\subsection{Calibration of the model}
In order to calibrate the above model, we can measure the empirical response function
%correlation between the order flow and price changes (we can also call it the price response function)
$\mathcal{R}(\ell)=\mathbb{E}[(m_{t+\ell}-m_t) \cdot \epsilon_t]$ and the empirical correlation function of the order signs $C(\ell)=\mathbb{E}[\epsilon_t\epsilon_{t+\ell}]$. These two functions form a linear system of equations
\begin{equation*}
  \mathcal{R}(\ell)=\sum_{0<n\leq\ell}G(n)C(\ell-n)+\sum_{n>0}\left[G(n+\ell)-G(n)\right]C(n),
\end{equation*}
whose solution is the propagator function $G(\ell)$, for $\ell>0$. 

An alternative method of estimation, which is less sensitive to boundary effects, uses the return process of Eq. \ref{eqn:prom_return_process}, such that the associated response function $\mathcal{S}(\ell)=\mathbb{E}[r_{t+\ell}\cdot\epsilon_t]$ and $C(\ell)$ are related through:
\begin{equation*}
  \mathcal{S}(\ell)=\sum_{n \geq 0}{\cal G}(n)C(n-\ell),
%\label{eqn:transient_estimation}
\end{equation*}
whose solution represents the values of the kernel ${\cal G}(\ell)$. The relation between $\mathcal{R}(\ell)$ and $\mathcal{S}(\ell)$ is:
\begin{equation}
  \mathcal{R}(\ell)=\sum_{0\leq i < \ell} \mathcal{S}(i)
  \label{eqn:positivelag}
\end{equation}
allowing to recover the response function from its differential form.

Once the propagator $G(\ell)$ is calibrated on the data, the model is fully specified by the statistics of the noise $\eta_t$. For simplicity, we will assume that $\eta_t$ has a low-frequency, white noise part of variance $D_\mathrm{LF}$, describing any ``news'' component not captured by the order flow itself, and a fast mean-reverting component of variance $D_\mathrm{HF}$ describing e.g. high frequency activity inside the spread (affecting the position of the mid-point $m_t$) or possible errors in the data itself. 

\subsection{Direct tests of the model}
Once the model is fully calibrated on data, we examine its performance by considering the prediction of two quantities, namely the negative lag response function and the signature plot. The former is the extension of the price response function, $\mathcal{R}(\ell)$, to $\ell<0$ values, measuring the correlation between the present sign of the market order and the past price changes:
\begin{equation}
  \mathcal{R}(-\ell)=-\sum_{0 < i \leq \ell} \mathcal{S}(-i)= - \mathbb{E}[(m_t - m_{t-\ell}) \cdot \epsilon_t].
  \label{eqn:negativelag}
\end{equation}
$\mathcal{R}(-\ell)$, with $\ell > 0$, is fully specified by the model, independently of $D_\mathrm{LF}$ and $D_\mathrm{HF}$. Naturally the one propagator model assumes a ``rigid'' order flow that does not adapt to price changes and leads to:
\begin{equation}
  \mathcal{R}^{\mathrm{TIM1}}(-\ell)= - \sum_{0<i\leq \ell} \sum_{n \geq 0} \mathcal{G}(n)C(n+i) < 0.
  \label{eqn:negativelag_prop}
\end{equation}
where TIM1 reminds us that this is the prediction according to the one propagator model. 
Empirically, however one expects that the order flow should be adapting to past price changes, and an upward movement of the price should attract more sellers (and vice-versa). In section \ref{sec:prop_empirical_test} we will compare the prediction of Eq.~\eqref{eqn:negativelag_prop} to empirical results.

The second prediction of the propagator model concerns the scale-dependent volatility of price changes, or ``signature plot'', defined as:
\begin{equation*}
  D(\ell) = \frac{1}{\ell} \mathbb{E}[(m_{t+\ell} - m_t)^2]. 
\end{equation*}
Using the propagator model, one finds the following exact expression:
\begin{equation*}
  D^\mathrm{TIM1}(\ell) = \frac{1}{\ell} \sum_{0 \leq n < \ell} G^2(\ell-n) + \frac{1}{\ell} \sum_{n > 0} \left[G(\ell+n)-G(n)\right]^2 + 2 \Psi(\ell)+ \frac{D_\mathrm{HF}}{\ell} + D_\mathrm{LF},
\end{equation*}
where $\Psi(\ell)$ is the correlation-induced contribution to the price diffusion:
\begin{align*}
  \ell \Psi(\ell) &= \sum_{0 \leq n < m < \ell} G(\ell-n) G(\ell-m) C(m-n) \\ 
  &+ \sum_{0 \leq n < m} \left[G(\ell+n)-G(n)\right]\left[G(\ell+m)-G(m)\right]C(m-n) \\
  &+ \sum_{0 \leq n < \ell} \sum_{m > 0} G(\ell-n) \left[G(\ell+m)-G(m)\right] C(m+n).
\end{align*}

Hence, once $G(\ell)$ is known, the signature plot of the price process can be computed and compared with empirical data. 

\subsection{Transient impact vs. history dependent impact}
The above model describes trades that impact prices, but with a time dependent, decaying impact function $G(\ell)$. One can in fact interpret the same model slightly differently, by writing as an identity:
\begin{equation}
  r_t= G(1)(\epsilon_t - \widehat{\epsilon}_t)+ \eta_t, \qquad \widehat{\epsilon}_t= - \sum_{\ell > 0} \frac{{\cal G}(\ell)}{G(1)}\epsilon_{t-\ell}.
  \label{eqn:hdim_return_process}
\end{equation}
This can be read as a model where the deviation of the realized sign $\epsilon_t$ from an expected level $\widehat{\epsilon}_t$ impacts the price linearly and permanently. If $\widehat{\epsilon}_t$ is the best possible predictor of $\epsilon_t$, then the above equation leads by construction to an exact martingale for the price process (i.e. the conditional average of $r_t$ on all past information is zero) \cite{mrr}. Since the impact depends on the past history of order flow, following Ref. \cite{eisler2012b}, we refer to the model on the left of Eq. \ref{eqn:hdim_return_process} as the History Dependent Impact Model and since only one type of past events is considered in the predictor we label it with HDIM1. When the best predictor is furthermore {\it linear} in the past order signs (as in the right equation of Eq. \ref{eqn:hdim_return_process}), then the TIM1 defined by Eq. \ref{eqn:prom_return_process} is {\it equivalent} to the HDIM1, Eq. \ref{eqn:hdim_return_process}. We will see below that as soon as one attempts to generalize the propagator model to multiple event types, TIM and HDIM become no longer equivalent. 

\subsection{The DAR process for trade signs}
\label{sec:dar_model}
When is the best predictor of the future price a linear combination of past signs, such that TIM and HDIM are equivalent when restricted to one type of market orders only? The answer is that this is true whenever the string of signs is generated by a so-called Discrete Autoregressive (DAR) process (see \cite{jacobs1978}). DAR processes are constructed as follows (our description here lays the ground for the more general MTD models described in the companion paper). The sign at time $t$ is thought of as the ``child'' of a previous sign $t - \ell$, where the distance $\ell$ is a random variable distributed according to a certain discrete distribution $\lambda_\ell$, with:
\begin{equation*}
  \sum_{\ell=1}^\infty \lambda_\ell = 1.
\end{equation*}
If $\lambda_{\ell > p} \equiv 0$, the model is called as DAR(p), and involves only $p$ lags. Once the ``father'' sign is chosen, one postulates that:
\begin{align*}
  \epsilon_t = \epsilon_{t-\ell} &\qquad \mbox{with probability~} \rho \\
  \epsilon_t = -\epsilon_{t-\ell} &\qquad \mbox{with probability~} 1-\rho.
\end{align*}

One can then show that in the stationary state, the signs $\pm$ are equiprobable, and the sign auto-correlation function $C(\ell)$ obeys the following Yule-Walker equation:
\begin{equation*}
  C(\ell) = (2 \rho - 1) \sum_{n=1}^\infty \lambda_n C(\ell-n).
\end{equation*}

There is therefore a one-to-one relation between $\lambda_\ell$ and $C(\ell)$. Note that in the empirical case where $C(\ell)$ decays as a power-law $\ell^{-\gamma}$ with exponent $\gamma < 1$, one can show that $\lambda_\ell \sim \ell^{(\gamma-3)/2}$ and $\rho \to 1^-$.

Now, from the very construction of the process, the conditional average of $\epsilon_t$ is given by:
\begin{equation*}
  \widehat{\epsilon}_t = (2 \rho - 1) \sum_{\ell=1}^\infty \lambda_\ell \epsilon_{t-\ell},
\end{equation*}
such that one can indeed identify the HDIM1 with a TIM1, with:
\begin{equation*}
  {\cal G}(\ell) = - (2 \rho - 1) G(1) \lambda_\ell.
\end{equation*}
When $C(\ell) \stackrel{{\ell \gg 1}}{\sim} \ell^{-\gamma}$, one finds as expected $G(\ell)=G(1) + \sum_{n=1}^\ell {\cal G}(n) \stackrel{{\ell \gg 1}}{\sim} \ell^{-\beta}$ with $\beta = (1- \gamma)/2$ \cite{bouchaud2004}.

\section{An extended propagator model with two types of market orders}
In order to develop the idea that large market orders (compared to the volume at the opposite best) may have a different impact than small ones, we need to extend the above propagator model to different events $\pi_t$, where we choose here two types of events $\pi_t$ defined as:
\begin{equation*}
  \pi_t =
  \left\{
    \begin{array}{ll}
      \mathrm{NC} & \mbox{if } r_t= m_{t+1} - m_t = 0 \\
      \mathrm{C} & \mbox{if }  r_t = m_{t+1} - m_t \neq 0.
    \end{array}
  \right.
\end{equation*}

We follow the general framework of \cite{eisler2012a}, but here the definition of price changing events is different. They refer to the total returns until the next transaction and they include the behaviour of liquidity takers and liquidity providers. These different events are discriminated by using indicator variables denoted as $I(\pi_t=\pi)$. The indicator, $I(\pi_t=\pi)$, is $1$ if the event at $t$ is of type $\pi$ and zero otherwise. The time average of the indicator function is the unconditional probability of event $\pi$, $\mathbb{P}(\pi)=\mathbb{E}[I(\pi_t=\pi)]$. The usage of the indicator function simplifies the calculation of the conditional expectations, which will be intensively used in the following. For example, if a quantity $X_{\pi_t,t}$ depends on the event type $\pi$ and the time $t$, then its conditional expectation is
\begin{equation*}
  \mathbb{E}[X_{\pi_t,t}|\pi_t=\pi]=\frac{\mathbb{E}[X_{\pi_t,t}I(\pi_t=\pi)]}{\mathbb{P}(\pi)}.
\end{equation*}
By definition of the indicator function we have that
\begin{equation*}
  \sum_\pi I(\pi_t=\pi)=1; \qquad \mbox{and} \qquad \sum_\pi X_{\pi,t}I(\pi_t=\pi)=X_{\pi_t,t}.
\end{equation*}

\subsection{Generalisation of the TIM}
At this stage, the natural generalisation of the TIM is to write the return process as 
\begin{equation*}
  r_t=\sum_\pi G_\pi(1) I(\pi_{t}=\pi) \epsilon_{t} + \sum_{t' < t} \sum_{\pi'}{\cal G}_{\pi'}(t-t')I(\pi_{t'}=\pi')\epsilon_{t'}+\eta_t, \quad {\cal G}_{\pi'}(\ell)\equiv G_{\pi'}(\ell+1)-G_{\pi'}(\ell),
  %\label{eqn:two_prom_return_process}
\end{equation*}
where $\pi=\{\mathrm{NC},\mathrm{C}\}$. Therefore we call this model TIM2. The resulting price process is a linear superposition of the decaying impact of different (signed) events:
\begin{equation}
  m_t=\sum_{t' < t}\left[\sum_{\pi} G_\pi(t-t')I(\pi_{t'}=\pi)\epsilon_{t'}+\eta_{t'}\right]+m_{-\infty}.
  \label{eqn:two_prom_price_process}
\end{equation}
which can be used to compute the signature plot $D(\ell)$ of model (see Appendix \ref{app:tim_signature_plot}).

The TIM2 can be calibrated very similarly as the TIM1 above, by noting that the differential response function
\begin{equation*}
  \mathcal{S}_\pi(\ell)=\mathbb{E}[r_{t+\ell}\cdot \epsilon_t|\pi_t=\pi]=\frac{\mathbb{E}[r_{t+\ell}\cdot\epsilon_t I(\pi_t=\pi)]}{\mathbb{P}(\pi)},
\end{equation*}
and the conditional correlation\footnote{It should be noted that $ C_{\pi_1,\pi_2}(\ell)$ is not bounded in $[-1,1]$ because we normalize the expectation in the numerator by the product $\mathbb{P}(\pi_1)\mathbb{P}(\pi_2)$ rather than by the joint probability $\mathbb{P}(\pi_t=\pi_1,\pi_{t+\ell}=\pi_2)$. This choice is done for speeding the computations and we have verified that the difference is very small.} of order signs of a pair of events $\pi_1$ and $\pi_2$
%\textcolor{red}{BENCE: the following equation is not correct: the middle term is not the same as the RHS. Why do we use the middle term at all? In all papers we use the correlation that is in the numerator and just "normalise" it by \mathbb{P}(pi1)\mathbb{P}(pi2). This way we get a measure that also gives info about the increased probability of an event occurring, not only the bias in the signs.}
\begin{equation}
  C_{\pi_1,\pi_2}(\ell)=\frac{\mathbb{E}[\epsilon_t I(\pi_t=\pi_1) \cdot \epsilon_{t+\ell} I(\pi_{t+\ell}=\pi_2)]}{\mathbb{P}(\pi_1)\mathbb{P}(\pi_2)}
  \label{eqn:conditional_corr_order_signs}
\end{equation}
are related through:
\begin{equation}
  S_{\pi_1}(\ell)=\sum_{\pi_2}\mathbb{P}(\pi_2)\sum_{n \geq 0}{\cal G}_{\pi_2}(n)C_{\pi_1,\pi_2}(\ell-n).
  \label{eqn:extended_transient_estimation}
\end{equation}

We use these quantities to evaluate the conditional response function ${\cal R}_\pi(\ell)=\sum_{i\le 0<\ell} S_\pi(\ell)$, the total impact function $\mathcal{S}(\ell)=\sum_\pi \mathbb{P}(\pi) \mathcal{S}_\pi(\ell)$ and the corresponding response function $\mathcal{R}(\ell)$. As for the TIM1, once we have calibrated ${\cal G}_\pi(\ell)$, we compute the predicted values of these response functions for negative lags, $\mathcal{R}_\pi^{\mathrm{TIM2}}(\ell)$ and $\mathcal{R}^{\mathrm{TIM2}}(\ell)$, and the predicted signature plot $D^{\mathrm{TIM2}}(\ell)$.

\subsection{Generalisation of the HDIM} 
However, this is not the only generalisation of the propagator model. In fact, the HDIM formulation, Eq. \ref{eqn:hdim_return_process}, lends itself to the following, different extension:
\begin{equation*}
  r_t= \sum_\pi G_\pi(1) I(\pi_{t}=\pi) \epsilon_{t} + \sum_{t' < t} \sum_{\pi',\pi} {\kappa}_{\pi',\pi} (t-t') I(\pi_t=\pi) I(\pi_{t'}=\pi')\epsilon_{t'} + \eta_t,
  %\label{eqn:two_hdim_return_process}
\end{equation*}
meaning that the expected sign for an event of type $\pi$ is a linear regression of past signed events, with an ``influence kernel'' $\kappa$ that depends on both the past event type $\pi'$ and  the current event $\pi$. This model is the HDIM2. It is clear that TIMs are actually special cases of HDIMs, with the identification:
\begin{equation}
  \kappa_{\pi',\pi} (\ell) = {\cal G}_{\pi'}(\ell), \qquad \forall \pi,
  \label{eqn:hdim_tim_model}
\end{equation}
i.e. the influence kernel $\kappa$ does not depend on the present event type $\pi$: Only the type of the past event $\pi'$ matters. The calibration of this model turns out to be more subtle and is discussed in Appendix \ref{app:hdim_calibration} (where some errors and misprints appearing in the text of \cite{eisler2012b} 
are corrected).

As above, we may ask when it is justified to consider that the expected sign for an event of type $\pi$ is a linear regression of past signed events. This requires to generalize the DAR model described in section \ref{sec:dar_model} above to a multi-event framework. This will precisely be the aim of part II of this work, where we introduce MTDs as a natural generalisation of DAR for order book events.

\subsection{Tests of the two families of models} 
Much as for the simple propagator model, one can test the predictive power of the TIM and HDIM framework by comparing the conditional response functions for negative lags $\mathcal{R}_\pi(-\ell)=-\mathbb{E}[(m_{t}-m_{t-\ell})\cdot\epsilon_t|\pi_t=\pi]$, $\pi =\{\mathrm{NC},\mathrm{C}\}$  with empirical data, as well as the signature plot $D(\ell)$ of the price process. In the following section we will investigate the results of the estimation of the above models, and compare these predicted quantities with their empirical determination. Our conclusion, in a nutshell, is that introducing two types of events substantially increases the performance of the propagator models and that -- perhaps expectedly -- the HDIM fares better than TIM, but only very slightly.

\section{Empirical calibration}
\subsection{Dataset description}
We have analysed the trading activity of the 50 most traded stocks at NYSE and NASDAQ stock exchanges, during the period February 2013 - April 2013 with a total of 63 trading days. We have chosen a wide panel of stocks of different types in order to perform a deep analysis of the two markets. We have considered only the trading activity in the period 9:30-15:30 in all the days under analysis, in order to reduce intraday patterns of activity, such as volume traded, average spread, etc. In particular we try to avoid the trading activity just after the pre-auction and the closing period of the end of the trading day. After trimming the beginning and the end of each trading day, for each stock we concatenate the data on different trading days and carry out our analysis on these time series. The tick size of all the stocks is 0.01 USD.

In Table \ref{tab:data_stock} we list the details of the stocks analysed. In particular, we have listed the volatility in basis points, the average daily traded amount in USD, the average bid-ask spread in ticks, and the average tick size-price ratio and we ranked the stocks by these values. We can divide the sample in two different groups, which are the large and small tick stocks. The bid-ask spread of a large tick stock is most of the times equal to one tick, whereas small tick stocks have spreads that are typically a few ticks. We will emphasise in the following sections the very different behaviour of these two groups of stocks. There exist also a number of stocks in the intermediate region between large and small tick stocks, which have the characteristics of both types.

For the period studied, the stock of Apple Inc. (AAPL) had on average a bid-ask spread of $9.14$ ticks, clearly making it a small tick stock. On the other hand, Microsoft Inc. (MSFT), with average bid-ask spread being $1.00$ ticks is a good candidate for a large tick stock. To illustrate our empirical analysis, we chose to show results for these two stocks in the following.

\begin{table}[p]
\centering
\footnotesize
\begin{tabular}{lcclrlrlr}
\hline \hline 
\multicolumn{2}{c}{Average traded} & \multirow{2}{*}{Volatility (bp)} & \multicolumn{2}{c}{Average} & \multicolumn{2}{c}{Average tick size} \\
\multicolumn{2}{c}{volume (M\$)} & & \multicolumn{2}{c}{spread (tick)} & \multicolumn{2}{c}{price ratio} \\
\hline
AAPL & 1695.13 & 1.05 & PCLN & 38.40 & MU & 11.24 \\
FB & 935.17 & 1.86 & GOOG & 19.13 & BAC & 8.38 \\
GOOG & 764.73 & 1.58 & AAPL & 9.14 & INTC & 4.74 \\
MSFT & 451.80 & 1.21 & NFLX & 9.07 & CSCO & 4.71 \\
AMZN & 420.61 & 1.82 & AMZN & 8.39 & YHOO & 4.54 \\
TSLA & 373.28 & 7.20 & IDPH & 5.12 & GE & 4.31 \\
XOM & 337.04 & 0.95 & V & 2.78 & EMC & 4.12 \\
BAC & 324.83 & 2.20 & TSLA & 2.73 & FB & 3.59 \\
BEL & 304.55 & 1.28 & GS & 2.68 & GMZ & 3.58 \\
GILD & 294.27 & 1.83 & IBM & 2.55 & PFE & 3.58 \\
NFLX & 280.16 & 3.12 & BIDU & 2.48 & MSFT & 3.56 \\
C & 255.12 & 1.57 & CELG & 1.95 & ORCL & 2.93 \\
CSCO & 248.78 & 1.86 & BRK & 1.49 & WFC & 2.76 \\
PCLN & 247.75 & 3.18 & MMM & 1.42 & SBC & 2.76 \\
CMCSA & 241.75 & 1.78 & CHV & 1.36 & TSLA & 2.60 \\
GE & 239.93 & 1.65 & PM & 1.30 & KO & 2.57 \\
QCOM & 238.70 & 1.39 & BA & 1.27 & CMCSA & 2.50 \\
JNJ & 236.30 & 0.84 & SLB & 1.27 & GILD & 2.35 \\
EBAY & 227.06 & 1.69 & AMGN & 1.23 & MRK & 2.32 \\
CMB & 221.06 & 1.40 & XOM & 1.07 & C & 2.27 \\
INTC & 220.21 & 1.41 & WMT & 1.06 & BEL & 2.13 \\
CHV & 218.52 & 1.16 & HD & 1.05 & CMB & 2.04 \\
PFE & 217.23 & 1.37 & SBUX & 1.04 & EBAY & 1.85 \\
GMZ & 204.79 & 2.13 & PG & 1.02 & DIS & 1.80 \\
SBC & 204.01 & 1.20 & EBAY & 1.02 & SBUX & 1.77 \\
IBM & 203.51 & 1.14 & PEP & 1.02 & QCOM & 1.51 \\
PG & 202.61 & 1.03 & GILD & 1.02 & HD & 1.47 \\
WFC & 196.04 & 1.32 & QCOM & 1.02 & WMT & 1.38 \\
V & 195.18 & 1.40 & DIS & 1.01 & PEP & 1.31 \\
MU & 193.90 & 3.96 & JNJ & 1.00 & JNJ & 1.30 \\
YHOO & 187.33 & 2.15 & GMZ & 1.00 & SLB & 1.30 \\
BIDU & 185.04 & 3.03 & C & 1.00 & PG & 1.30 \\
KO & 172.95 & 1.38 & MRK & 1.00 & BA & 1.27 \\
DIS & 163.36 & 1.41 & CMB & 1.00 & AMGN & 1.13 \\
MRK & 161.24 & 1.57 & BEL & 1.00 & XOM & 1.12 \\
CELG & 157.84 & 2.04 & CMCSA & 1.00 & PM & 1.09 \\
IDPH & 151.86 & 3.33 & KO & 1.00 & BIDU & 1.09 \\
BRK & 151.23 & 1.75 & WFC & 1.00 & BRK & 0.99 \\
SBUX & 150.43 & 1.66 & ORCL & 1.00 & MMM & 0.96 \\
EMC & 146.03 & 1.84 & FB & 1.00 & CELG & 0.96 \\
AMGN & 141.99 & 1.72 & SBC & 1.00 & CHV & 0.85 \\
PEP & 140.86 & 1.05 & EMC & 1.00 & GS & 0.66 \\
WMT & 140.10 & 1.08 & PFE & 1.00 & V & 0.63 \\
BA & 137.41 & 1.65 & CSCO & 1.00 & IDPH & 0.60 \\
PM & 135.10 & 1.30 & YHOO & 1.00 & NFLX & 0.55 \\
SLB & 132.91 & 1.76 & INTC & 1.00 & IBM & 0.49 \\
GS & 130.26 & 1.81 & MSFT & 1.00 & AMZN & 0.38 \\
ORCL & 129.98 & 1.91 & GE & 1.00 & AAPL & 0.22 \\
HD & 128.21 & 1.62 & MU & 1.00 & PCLN & 0.14 \\
MMM & 125.19 & 1.47 & BAC & 1.00 & GOOG & 0.12 \\
\hline \hline 
\end{tabular}
\caption{Details of analysed stocks: rank by average traded daily amount (M\$), volatility, rank by average spread over tick size, and by average tick size (bp).}
\label{tab:data_stock}
\end{table}

\subsection{The one-event propagator model: calibration and tests}
\label{sec:prop_empirical_test}
\begin{figure}[p]
  \begin{minipage}[c]{0.5\columnwidth}
    \includegraphics[width=\columnwidth]{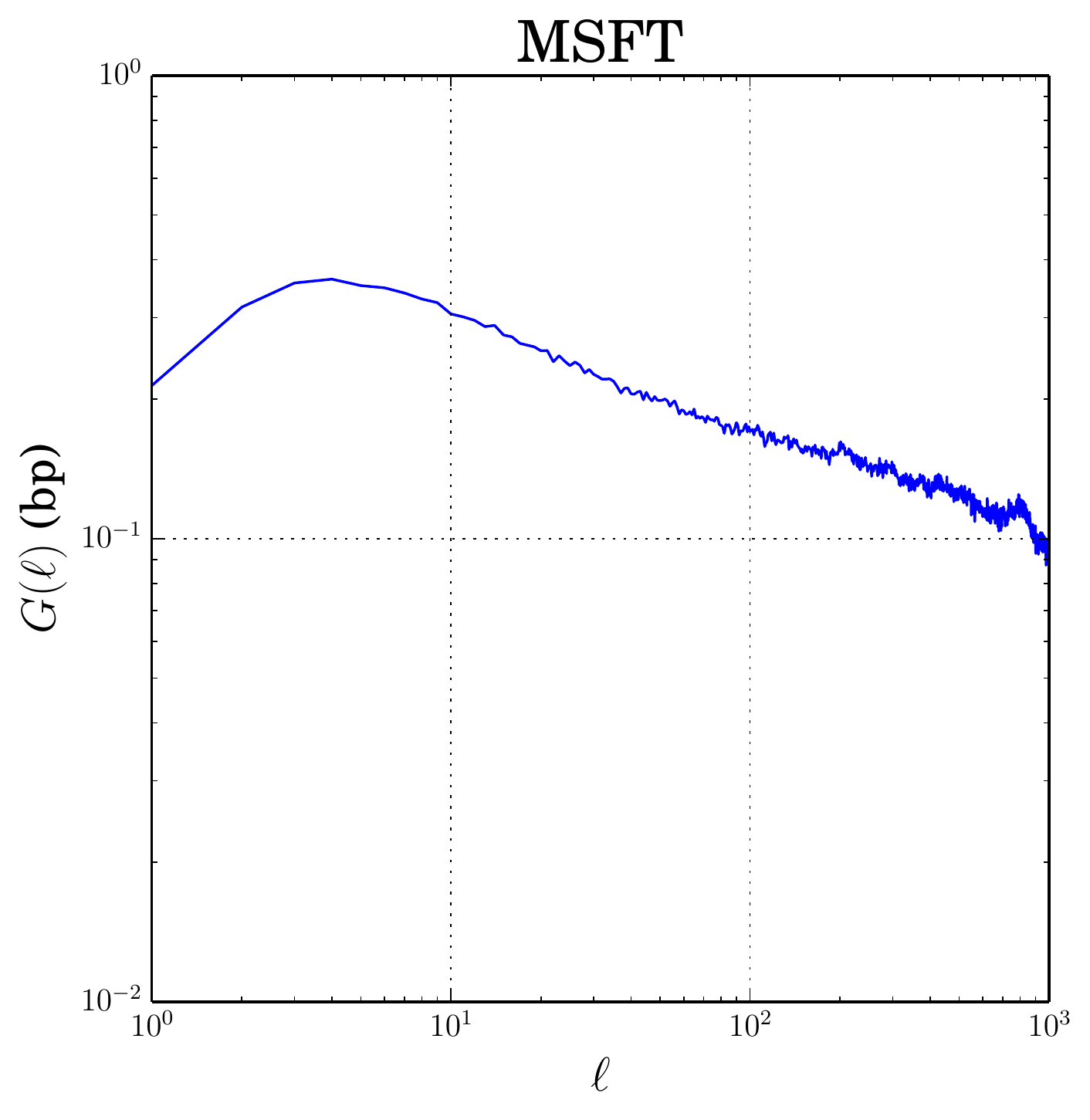}
    \vspace{5mm}
  \end{minipage}%
  \begin{minipage}[c]{0.5\columnwidth}
    \includegraphics[width=\columnwidth]{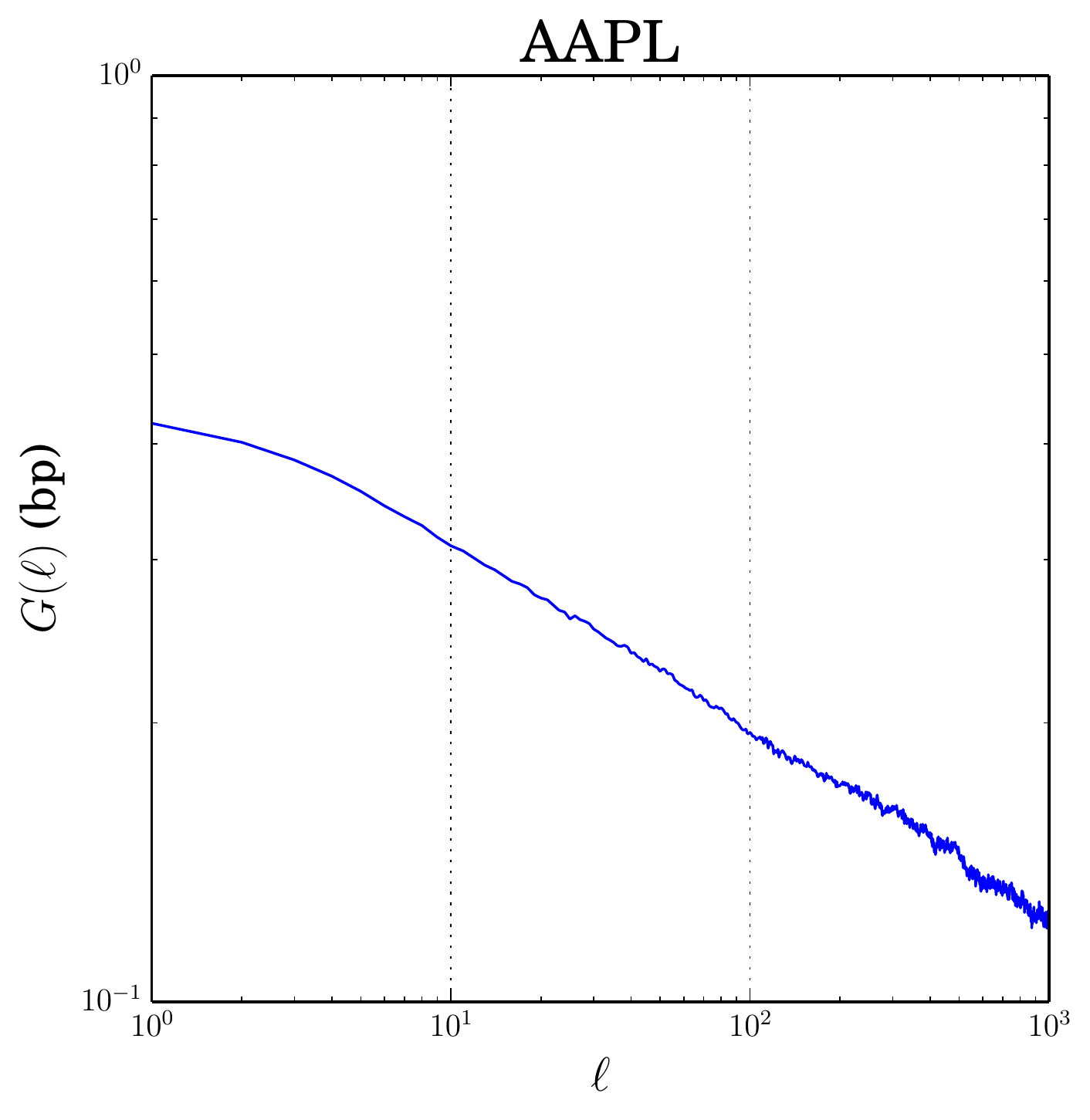}
    \vspace{5mm}
  \end{minipage}
  \begin{minipage}[c]{0.5\columnwidth}
    \includegraphics[width=\columnwidth]{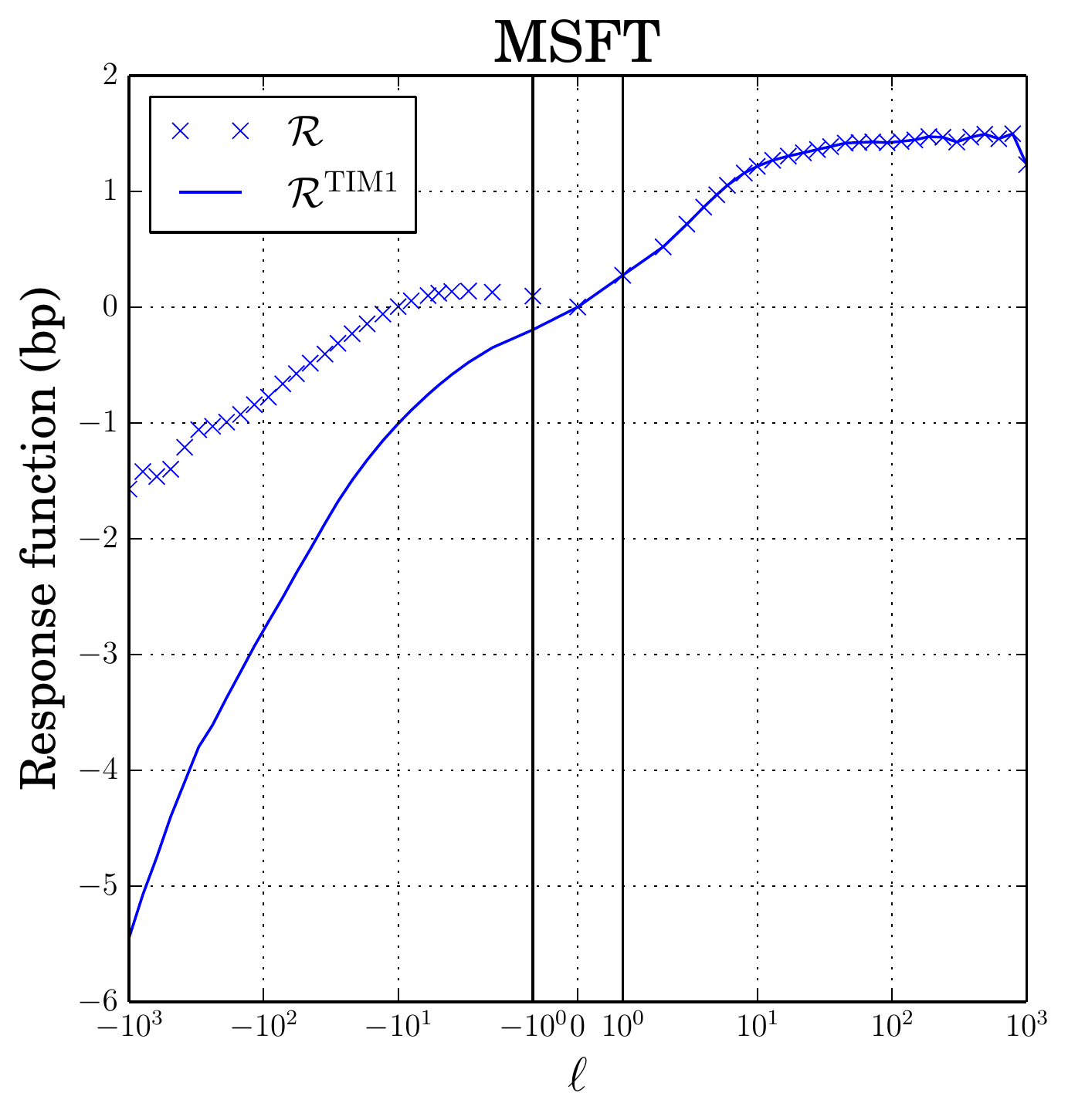}
    \vspace{5mm}
  \end{minipage}%
  \begin{minipage}[c]{0.5\columnwidth}
    \includegraphics[width=\columnwidth]{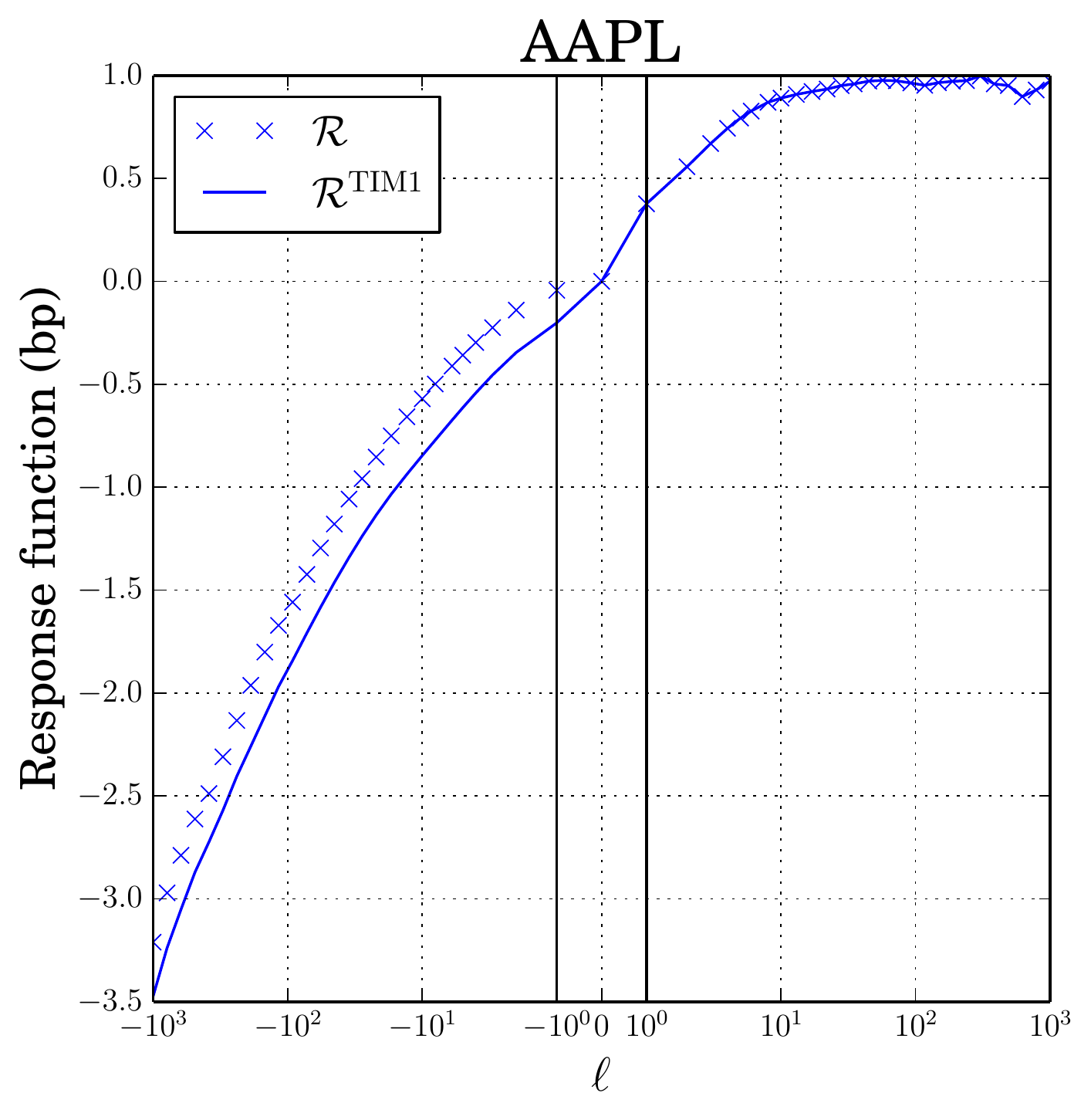}
    \vspace{5mm}
  \end{minipage}
  \caption{(Top panels) The estimated propagator functions for MSFT and AAPL. (Bottom panels) Response functions for positive and negative lags (blue markers) and the theoretical prediction of the estimated TIM1 (solid lines) for MSFT and AAPL. The scale for $\ell$ close to zero and bounded by the two vertical lines is linear, whereas outside this region the scale is logarithmic.}
  \label{fig:estimated_transient_impact_model}
\end{figure}

%The response function $\mathcal{R}(\ell)$ for positive $\ell$ is the average price change after the execution of a market order and it is an important measure of impact. This function (and its discrete derivative $\mathcal{S}(\ell)$) can be easily measured on real data together with the sign correlation function $C(\ell)$. We then determine the propagator $G(\ell)$ from Eq. \ref{eqn:transient_estimation}.
The top panels of Fig. \ref{fig:estimated_transient_impact_model} show the estimation of the propagators $G(\ell)$ for MSFT and AAPL. For both large and small tick stocks the decay of the propagator is slow, well above the noise level after 1000 transactions. We can see that for MSFT (as well as for other large tick stocks) the propagator function first increases for a few time lags, and starts decreasing only after that. Thus, the derivative ${\cal G}(\ell)$ is positive for small lags, and since $G(1)>0$ too, the market impact should be reinforced by a sequence of orders on the same side of the order book. This should lead to violations of the market efficiency on short time scales. This is a direct symptom of the inadequacy of the one-event propagator formalism for large ticks: in fact, we will see that the order flow cannot be considered to be independent of the price changes in this case. After an uptick move, there is a high probability that the next order will be in the opposite direction, reinstalling price efficiency. This will be well captured by the two-event propagator below. 

For AAPL and other small tick stocks we only see a monotone the decay of the propagator. The assumption of a rigid order flow, insensitive to price moves, will be approximately correct in that case (see \cite{toth2012}), the relaxation of the propagator alleviating the correlation of the signs. We can already anticipate that the two-event propagator framework will be much more beneficial for large tick stocks than for small tick stocks. 

\begin{figure}[t]
\centering
\includegraphics[width=0.9\columnwidth]{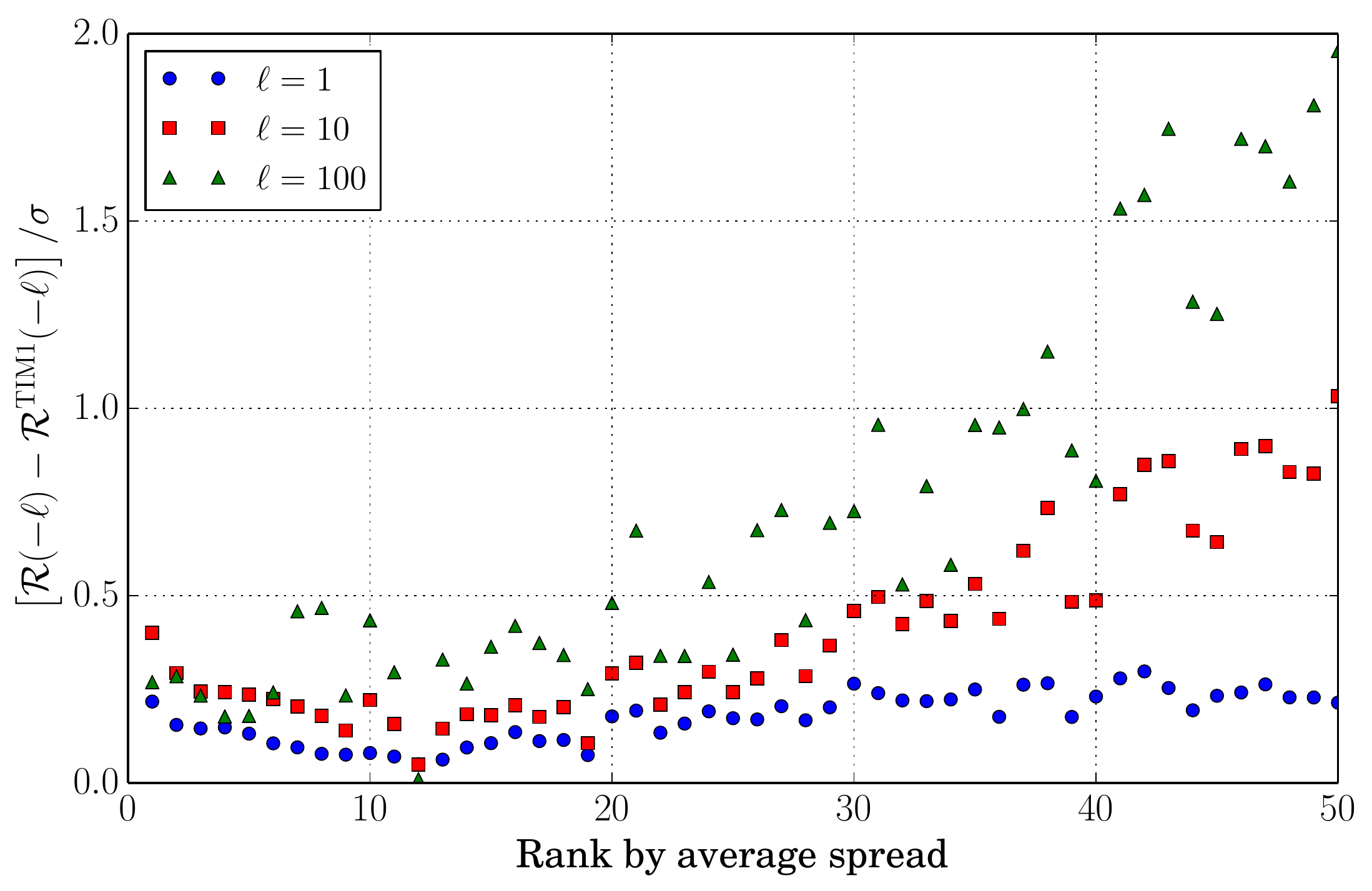}
\caption{Deviation from the TIM1 theoretical prediction of the response function at negative lags for 50 stocks under analysis ranked by the average spread. Small tick stocks are in the left side of the figure, whereas large tick stocks are in the right side}
\label{fig:anomalous_negative_stocks}
\end{figure}

The bottom panels of Fig. \ref{fig:estimated_transient_impact_model} show the price response for both positive and negative lags. The dashed lines in the plots show the theoretical prediction of the one-event propagator model by using the estimated kernels. In the case of MSFT the measured response function for negative lags $\ell$ is well above the prediction of the propagator model (solid line), that, as we discussed assumes a rigid order flow not depending on price changes. As anticipated above, this means that in the data there exists an additional anti-correlation between past returns and the subsequent order flow, which is not captured by the model. A similar, though much weaker deviation can be seen in the case of AAPL. In general, this effect is very pronounced in the case of large tick stocks, whereas in the case of small tick stocks  it exists but is much weaker. In fact, in Fig. \ref{fig:anomalous_negative_stocks} we plot the ratio $[\mathcal{R}(-\ell)-\mathcal{R}^{\text{TIM1}}(-\ell)]/\sigma$ for $\ell=1,10,100$, $\sigma$ being the volatility per trade, by ranking the stocks in the x-axis by the average spread. We observe that for small tick stocks (left part of the plot) the difference is relatively small, while for large tick stocks (right part of the plot) the prediction error on the negative lag response of the TIM1 becomes quite large, especially for large lags $\ell$.  
%Noting that $\mathcal{R}_\mathrm{th}(\ell<0)$ is negative, the quantity $z + 1$ is just the relative difference between the empirical response function and the theoretical prediction of the model with one propagator. As we can see, for every stock in our pool $z \geq -1$ meaning that a positive/negative return leads to an increased flow of sell/buy market orders. However, $z + 1$ is small for small tick stocks, in particular for $\ell = 100$, and steadily increases with the tick size. A one tick price move when the tick size is large is quite significant, and should indeed trigger market orders in the opposite direction when it happens. It also suggests, as will be clear below, that one should at the very least distinguish events of type $\pi =\{\mathrm{NC},\mathrm{C}\}$ in this case.

\begin{figure}[t]
  \begin{minipage}[c]{0.5\columnwidth}
    \includegraphics[width=\columnwidth]{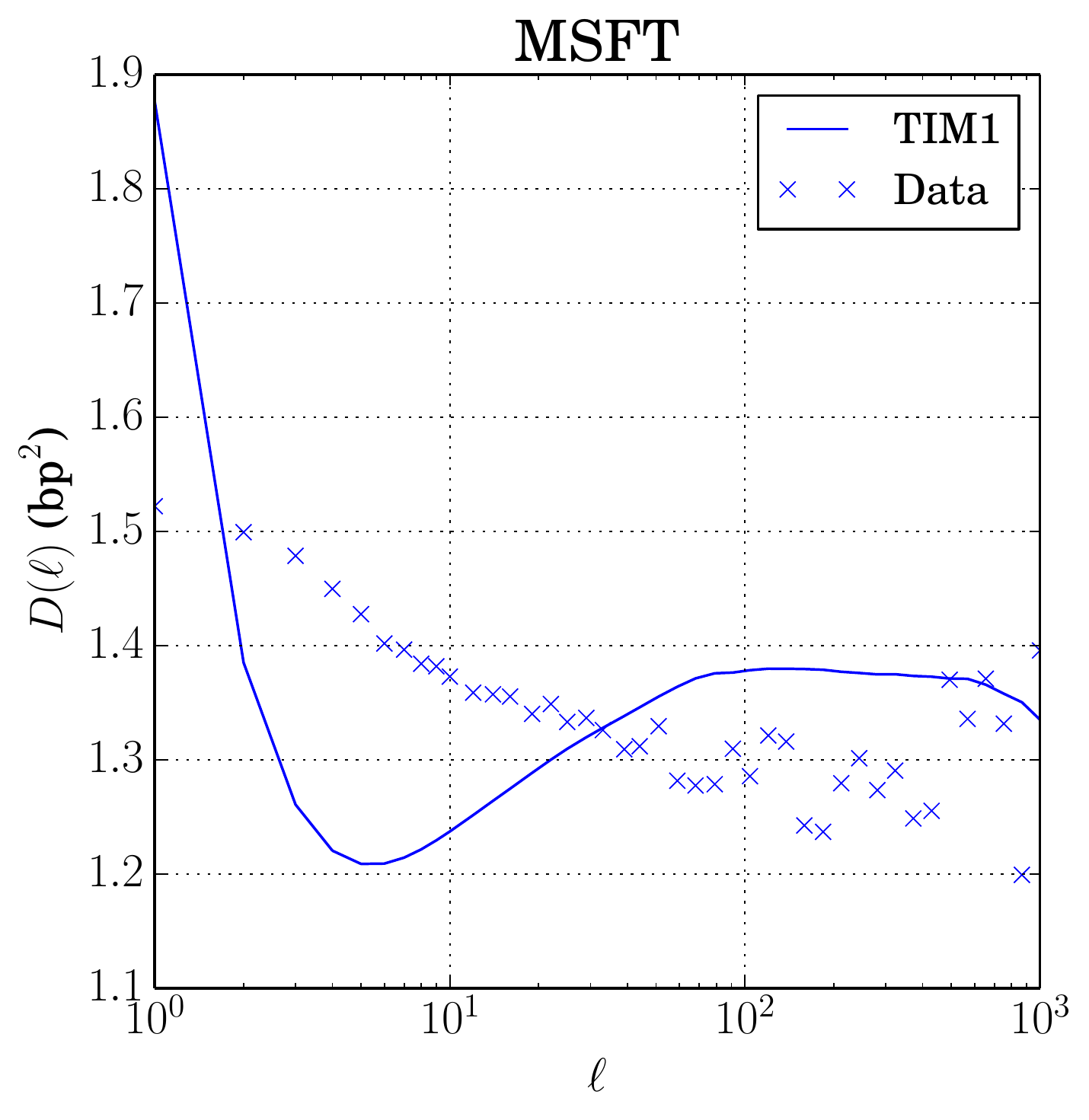}
    \vspace{5mm}
  \end{minipage}%
  \begin{minipage}[c]{0.5\columnwidth}
    \includegraphics[width=\columnwidth]{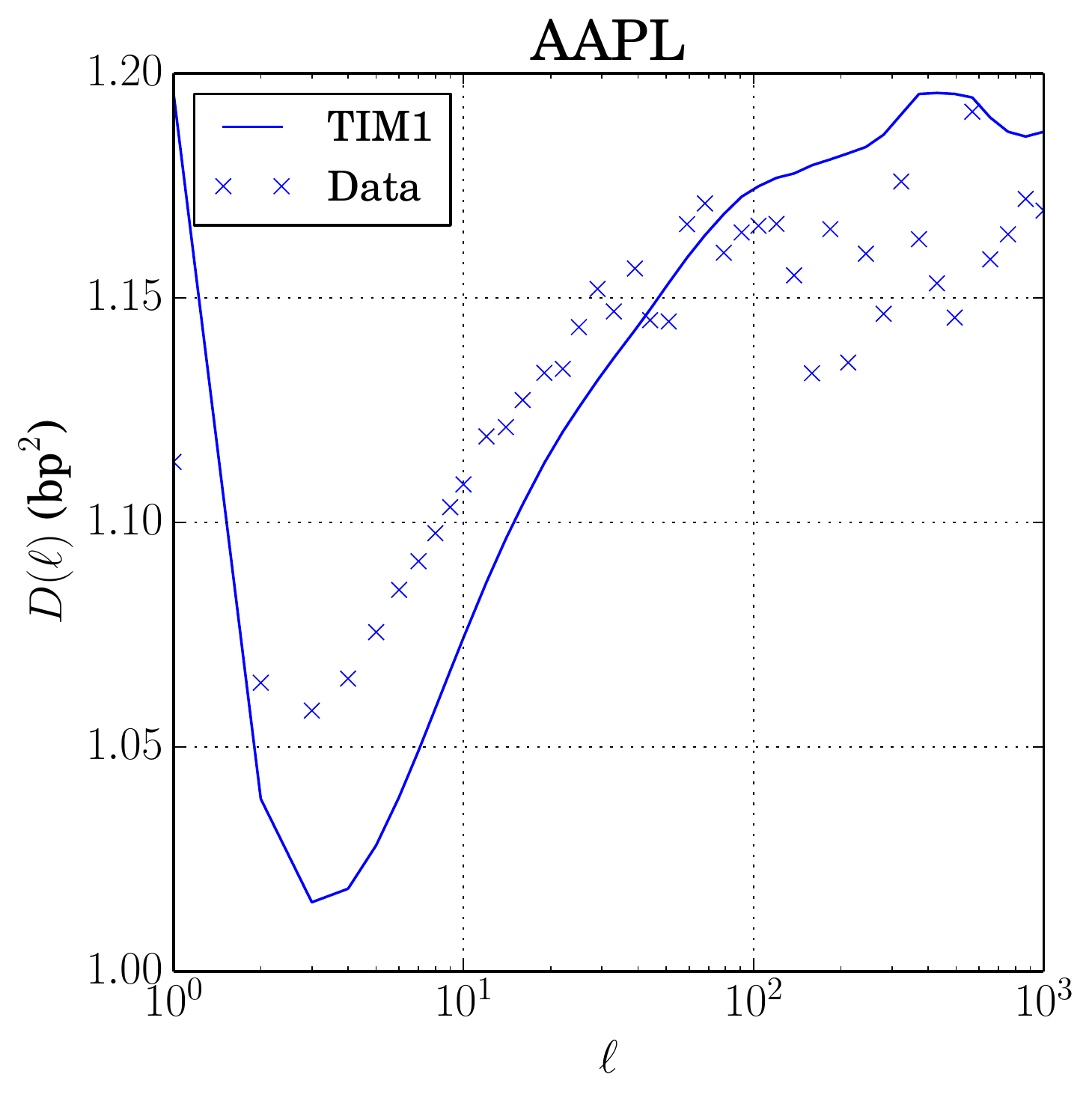}
    \vspace{5mm}
  \end{minipage}
  \caption{The empirical signature plot $D(\ell)$ and the theoretical curves of the estimated TIM1 for MSFT ($D_\mathrm{LF}=0.65$ and $D_\mathrm{HF}=1.13$) and APPL ($D_\mathrm{LF}=0.58$ and $D_\mathrm{HF}=0.46$).}
  \label{fig:estimated_transient_impact_model_signature_plot}
\end{figure}

Turning now to the signature plot $D(\ell)$, we see in Fig. \ref{fig:estimated_transient_impact_model_signature_plot} that small tick and large tick stocks behave very differently. For small tick stocks, we see that $D(\ell)$ increases with $\ell$ as soon as $\ell \geq 3$, corresponding to a ``trend-like'' behaviour. The decreasing behaviour of $D(\ell)$ for smaller lags corresponds to high frequency activity with the spread, leading to a minimum in $D(\ell)$. For large tick stocks this is absent and one finds ``mean-reverting'' behaviour, with a steadily decreasing signature plot. The prediction of the  one-event propagator model fares quite well at accounting for the trending behaviour of small tick stocks, provided the two extra fitting parameters $D_\mathrm{LF}$ and $D_\mathrm{HF}$ are optimized with OLS in order to minimize the distance between the empirical and the theoretical curves of the model. We note for example that choosing $D_\mathrm{LF}=0$ would underestimate (in the case of AAPL) the long-term volatility by a factor of two. For large tick stocks, however, the mean-reverting behaviour is completely missed. We now turn to propagator models that distinguish between price-changing and non price-changing market orders, and see how the situation for large tick stocks indeed greatly improves.

\subsection{Two-event propagator model}
The aim of this section is to show that an extended propagator model allows us to reproduce satisfactorily the additional anti-correlations between past returns and subsequent order signs (revealed by the discrepancy between $\mathcal{R}(\ell<0)$ and $\mathcal{R}^{\mathrm{TIM1}}(\ell<0)$) by including an implicit coupling between past returns and order flow. We will also require that the signature plot $D(\ell)$ is correctly accounted for, in particular for large tick stocks. 

\begin{figure}[p]  
  \begin{minipage}[c]{0.5\columnwidth}
    \includegraphics[width=\columnwidth]{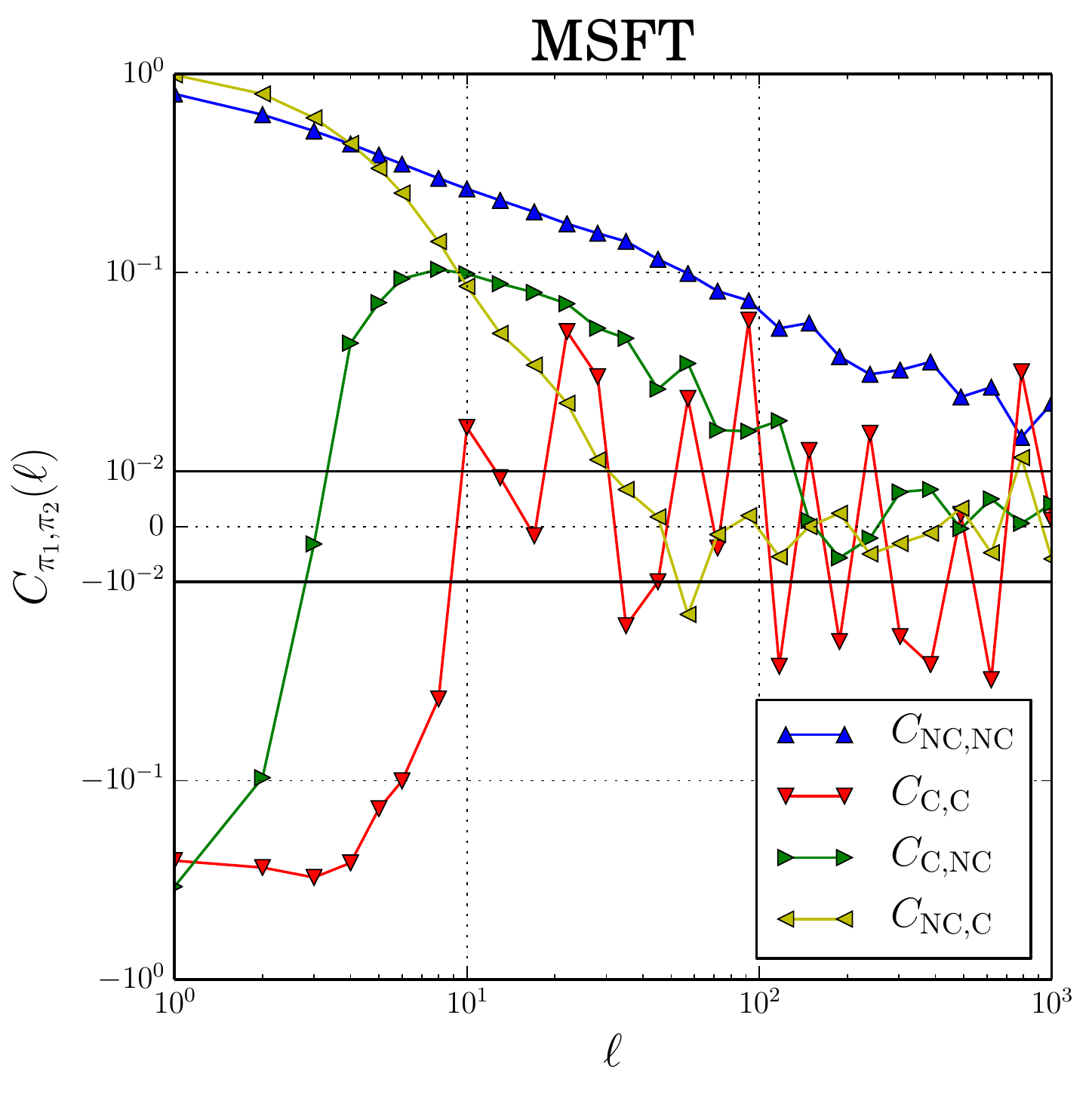}
    \vspace{5mm}
  \end{minipage}%
  \begin{minipage}[c]{0.5\columnwidth}
    \includegraphics[width=\columnwidth]{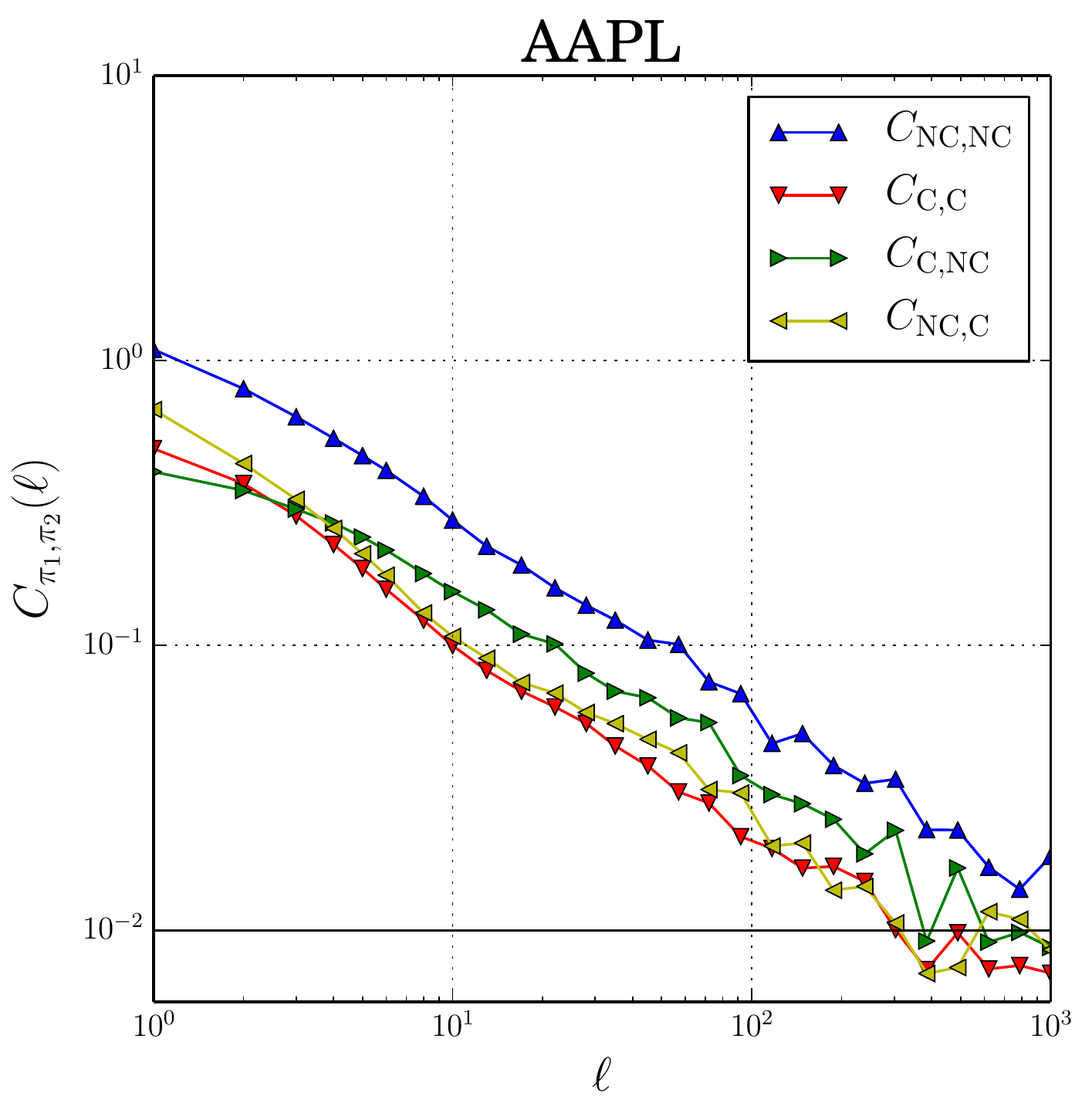}
    \vspace{5mm}
  \end{minipage} 
  \caption{Conditional correlations function of signed events  $C_{\pi_1,\pi_2}(\ell)$  measured on MSFT and APPL data. Note that the first subscript corresponds to the event that happened first chronologically. The scale for values of the correlations close to zero and bounded by horizontal solid lines is linear, whereas outside this region the scale is logarithmic.}
  \label{fig:estimated_cond_corr}
\end{figure}

\begin{figure}[p]    
  \begin{minipage}[c]{0.5\columnwidth}
    \includegraphics[width=\columnwidth]{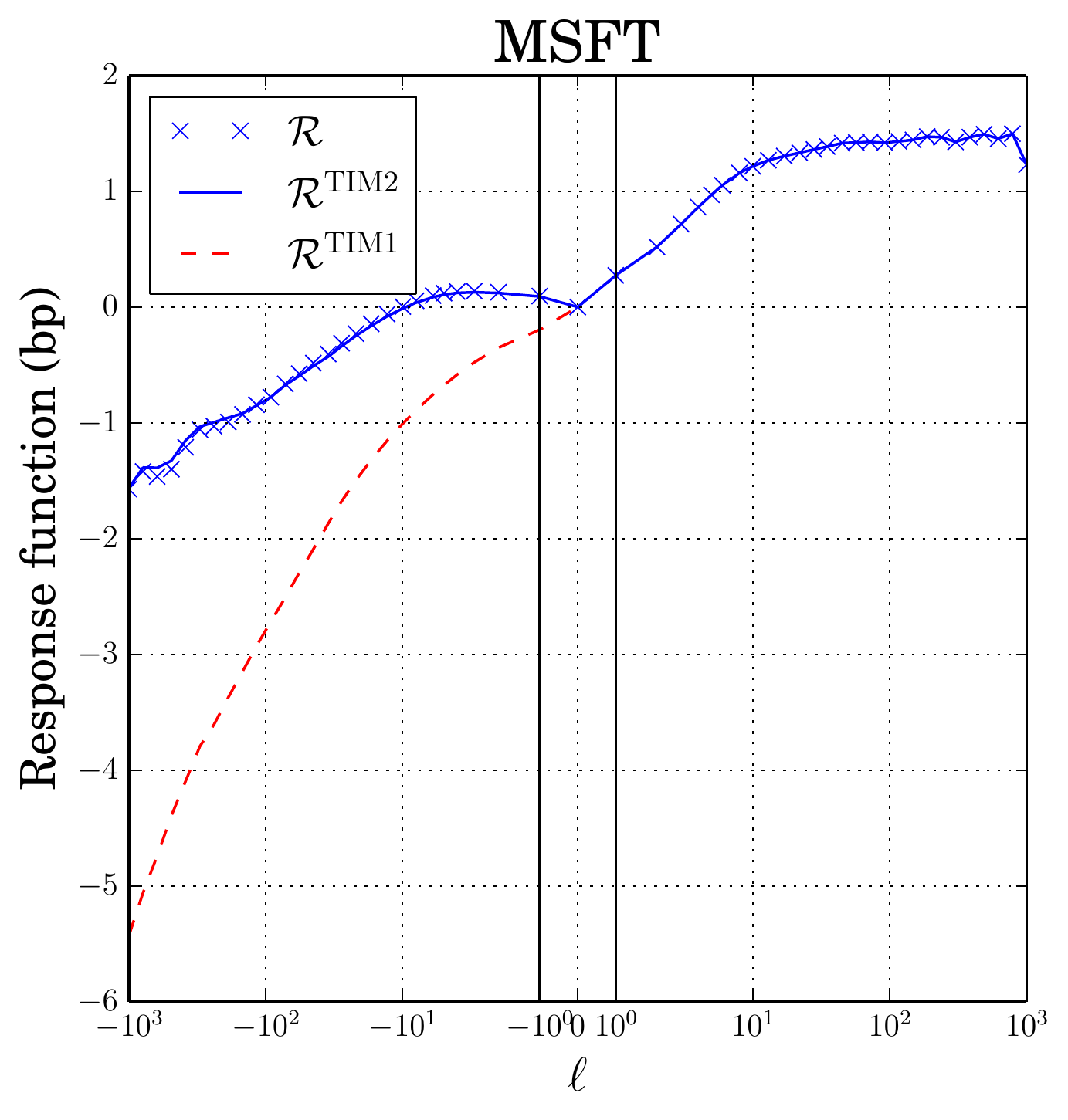}
    \vspace{5mm}
  \end{minipage}%
  \begin{minipage}[c]{0.5\columnwidth}
    \includegraphics[width=\columnwidth]{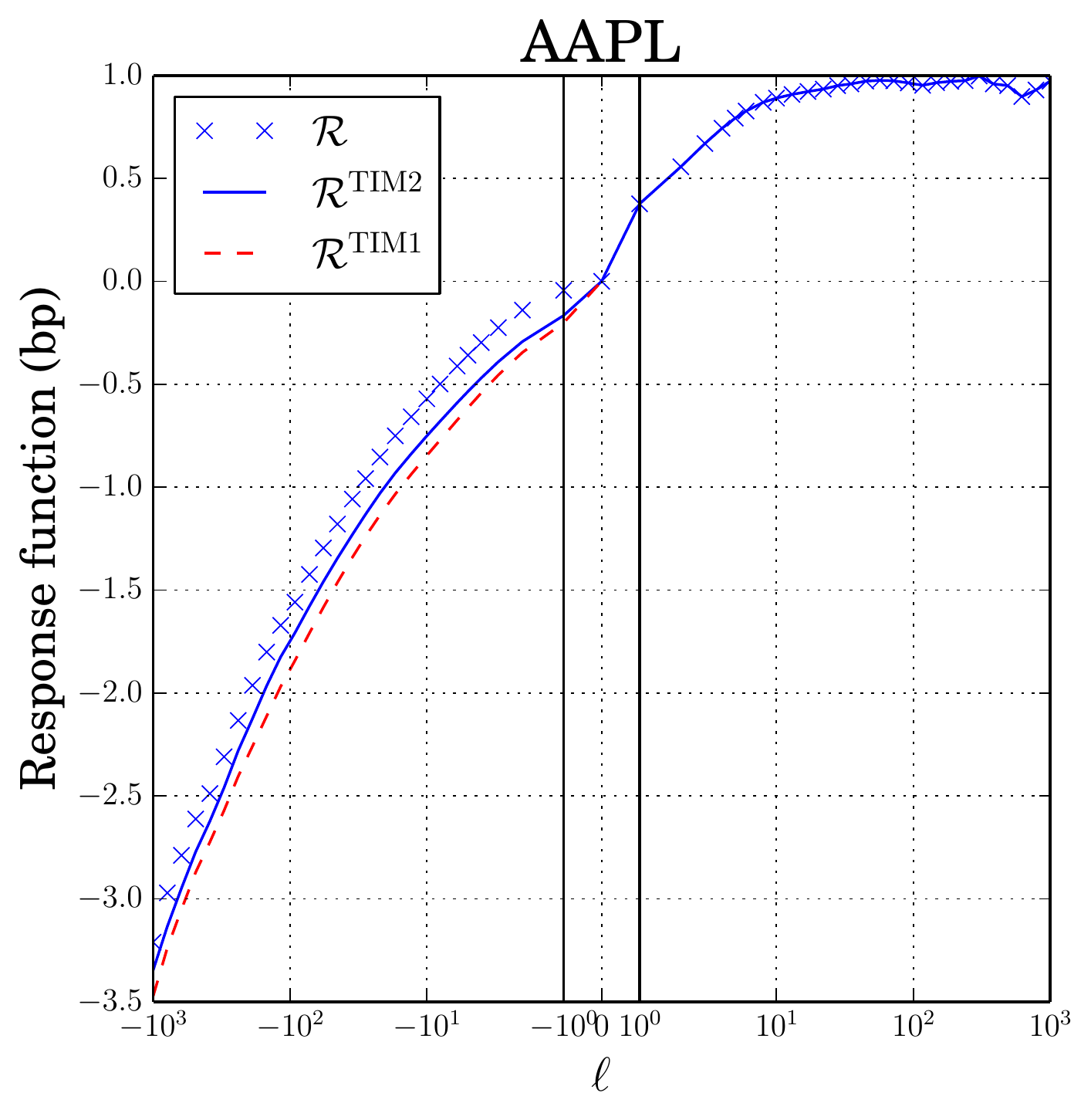}
    \vspace{5mm}
  \end{minipage}
  \caption{Conditional response function for positive and negative lags (blue markers) and the theoretical prediction of the TIM2 calibrated on MSFT and APPL (solid lines). Theoretical prediction of response function for negative lags for TIM1 (red dashed lines). The scale for $\ell$ close to zero and bounded by vertical solid lines is linear, whereas outside this region the scale is logarithmic.}
  \label{fig:estimated_response_function}
\end{figure}

The extended version of the propagator model with two events $\pi=\{\mathrm{NC},\mathrm{C}\}$ can follow two routes, as discussed above. One is the TIM2, which can be estimated much as the one-event model, by solving the linear system of Eq. \ref{eqn:extended_transient_estimation}. The second is the HDIM2, whose estimation involves determining the influence kernels $\kappa_{\pi_1,\pi_2} (\ell)$ for $\pi_2=\mathrm{C}$, because $\kappa_{\pi_1,\mathrm{NC}} (\ell)=0$ by construction. The calibration requires estimating three-point correlation functions or approximating them in terms of two-point correlations -- as detailed in Section~\ref{sec:hdim_estimation} we will follow the latter approximation. Thus, the correlation $C_{\pi_1,\pi_2}(\ell)$ of the different signed events, defined in Eq.~\eqref{eqn:conditional_corr_order_signs} is an important input of the calibration for both generalised linear models. Note that the first subscript corresponds to the event that happened first chronologically.  We start by showing its empirical estimation for the two typical stocks (see Fig. \ref{fig:estimated_cond_corr}). 

For AAPL, all auto-correlation and cross-correlation functions have almost the same power-law decay and they are all positive. This is expected since C and NC events are not radically different for small tick stocks. Note that the unconditional probability of price changing market orders is $\mathbb{P}(\pi=\mathrm{C})=0.69$. Correlation functions look similar for other small tick stocks too.

For MSFT the curves reveal a different behaviour. For example the $C_{\mathrm{NC},\mathrm{NC}}$ auto-correlation has the familiar power-law shape possibly due to order splitting. The $C_{\mathrm{NC},\mathrm{C}}$ correlation is also positive but decays faster. Note that it starts at $C_{\mathrm{NC},\mathrm{C}}(1)\approx 0.95$, which means that a C order immediately following a NC order is in the same direction with very high probability. This describes NC orders that leave a relatively small quantity at the best offer, which is then immediately ``eaten'' by the next market orders. Its relatively fast decay suggests that agents splitting their metaorders avoid being aggressive and nearly only send NC orders. The other two correlations $C_{\mathrm{C},\mathrm{C}}$ and $C_{\mathrm{C},\mathrm{NC}}$ both start negative and capture the effect we are interested in: After a price changing event, it is highly likely that the subsequent order flow (either C or NC) will be in the other direction. Note however that $\mathbb{P}(\pi=\mathrm{C})=0.08$ and that it is exceedingly rare to observe a succession of two C events separated by a small lag. This type of behaviour is the one that can be seen in general for large tick stocks.

\subsubsection{Tests on the TIM2}
The estimation procedure involves the empirical determination of the response function for positive lags, and allows us to calculate the theoretical prediction of the response function for negative lags, as well as the signature plot.

Fig. \ref{fig:estimated_response_function} shows the empirical response function for positive lags $\mathcal{R}(\ell>0)$ and negative lags $\mathcal{R}(\ell<0)$, together with the predicted response function $\mathcal{R}^\mathrm{TIM2}(\ell)$, according to the calibrated TIM2. In the case of large tick stocks the empirical curves are perfectly reproduced, whereas for small tick stocks some little deviation still persists. The improvement with respect to the TIM1 is quite remarkable. This can be seen the from comparison of the prediction of the response function for negative lags of the TIM1,  $\mathcal{R}^\mathrm{TIM1}(\ell<0)$, also plotted in Fig. \ref{fig:estimated_response_function}.

\begin{figure}[p]
  \begin{minipage}[c]{0.5\columnwidth}
    \includegraphics[width=\columnwidth]{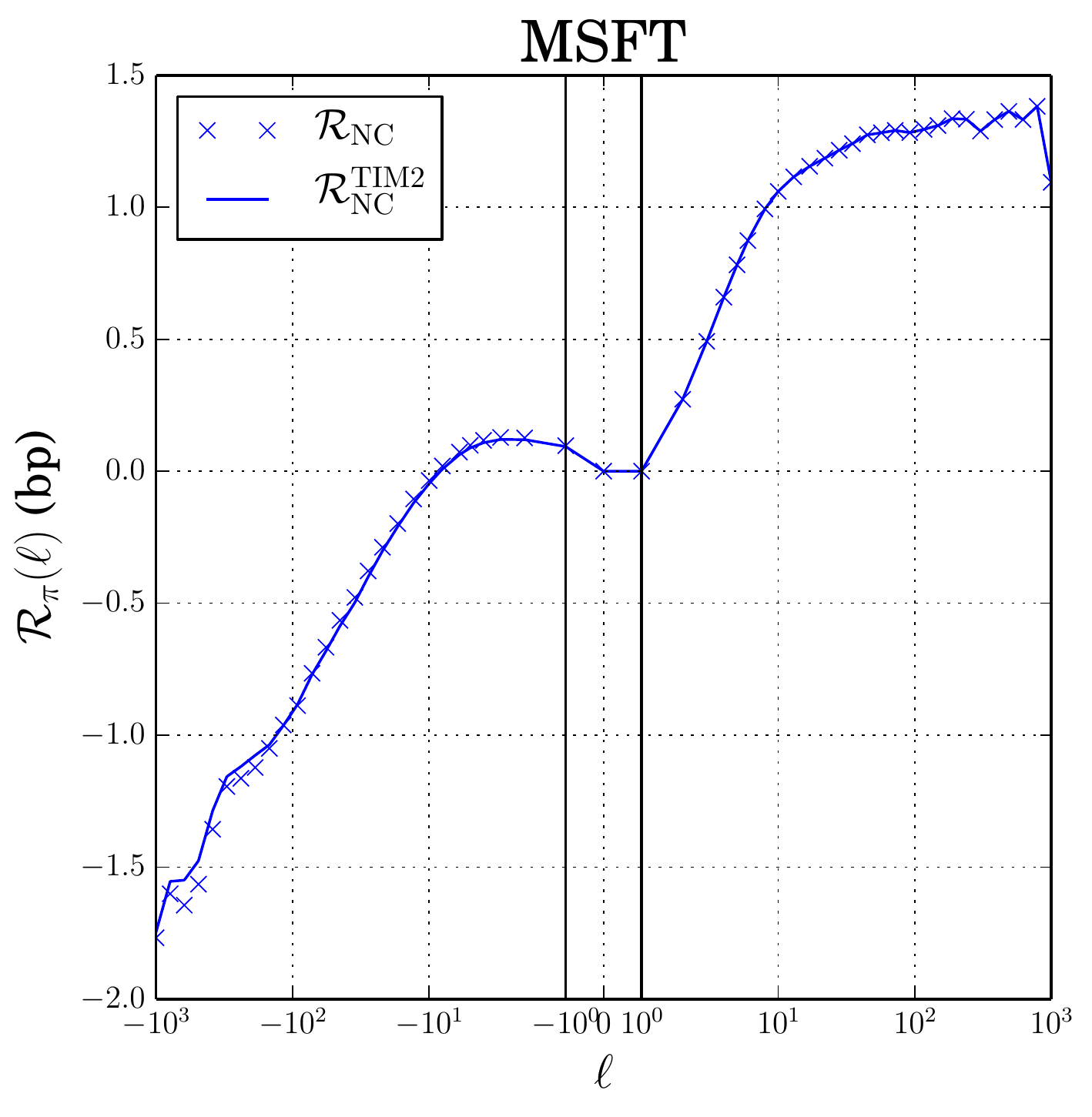}
    \vspace{5mm}
  \end{minipage}%
  \begin{minipage}[c]{0.5\columnwidth}
    \includegraphics[width=\columnwidth]{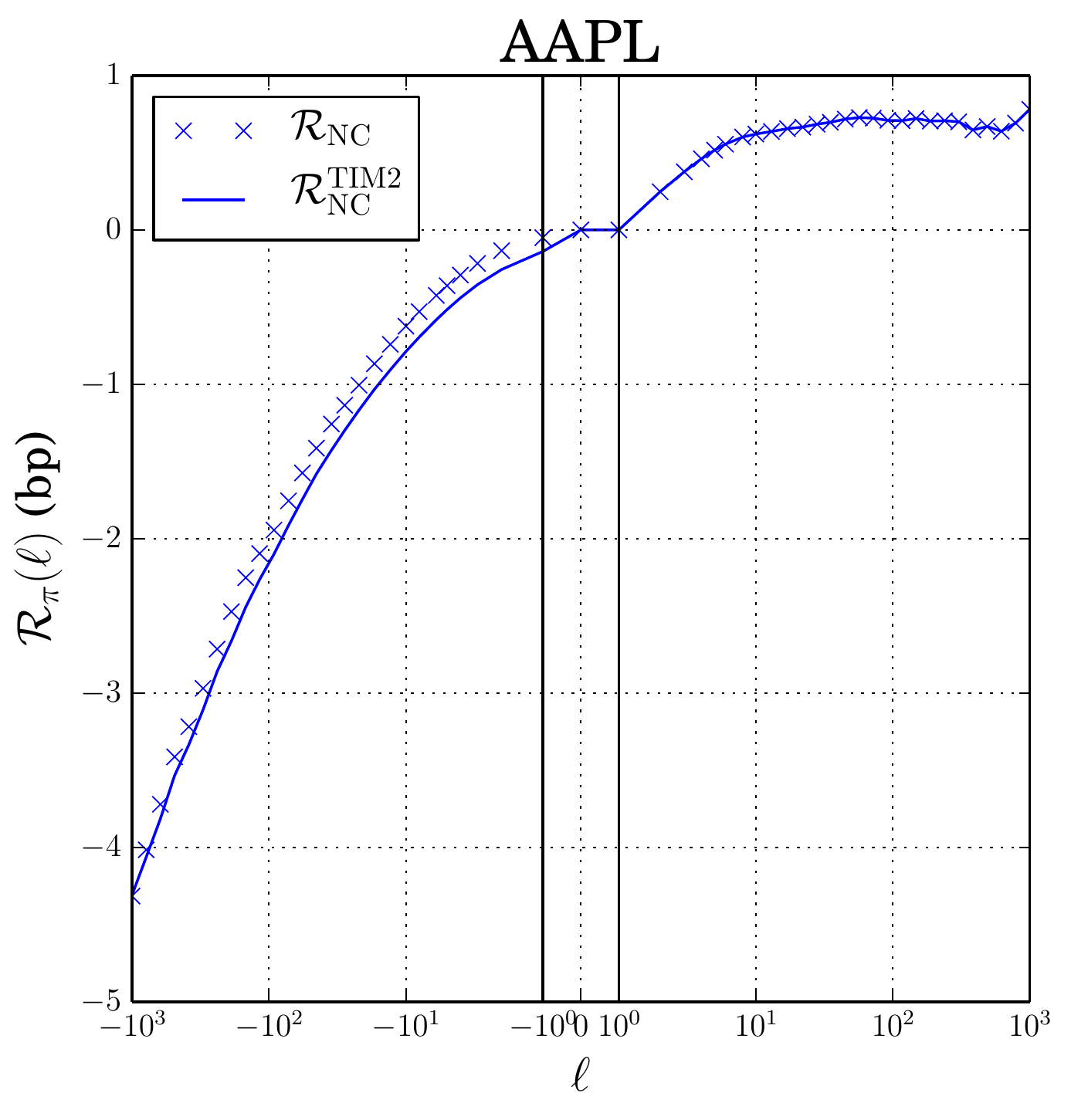}
    \vspace{5mm}
  \end{minipage} \\
  \begin{minipage}[c]{0.5\columnwidth}
    \includegraphics[width=\columnwidth]{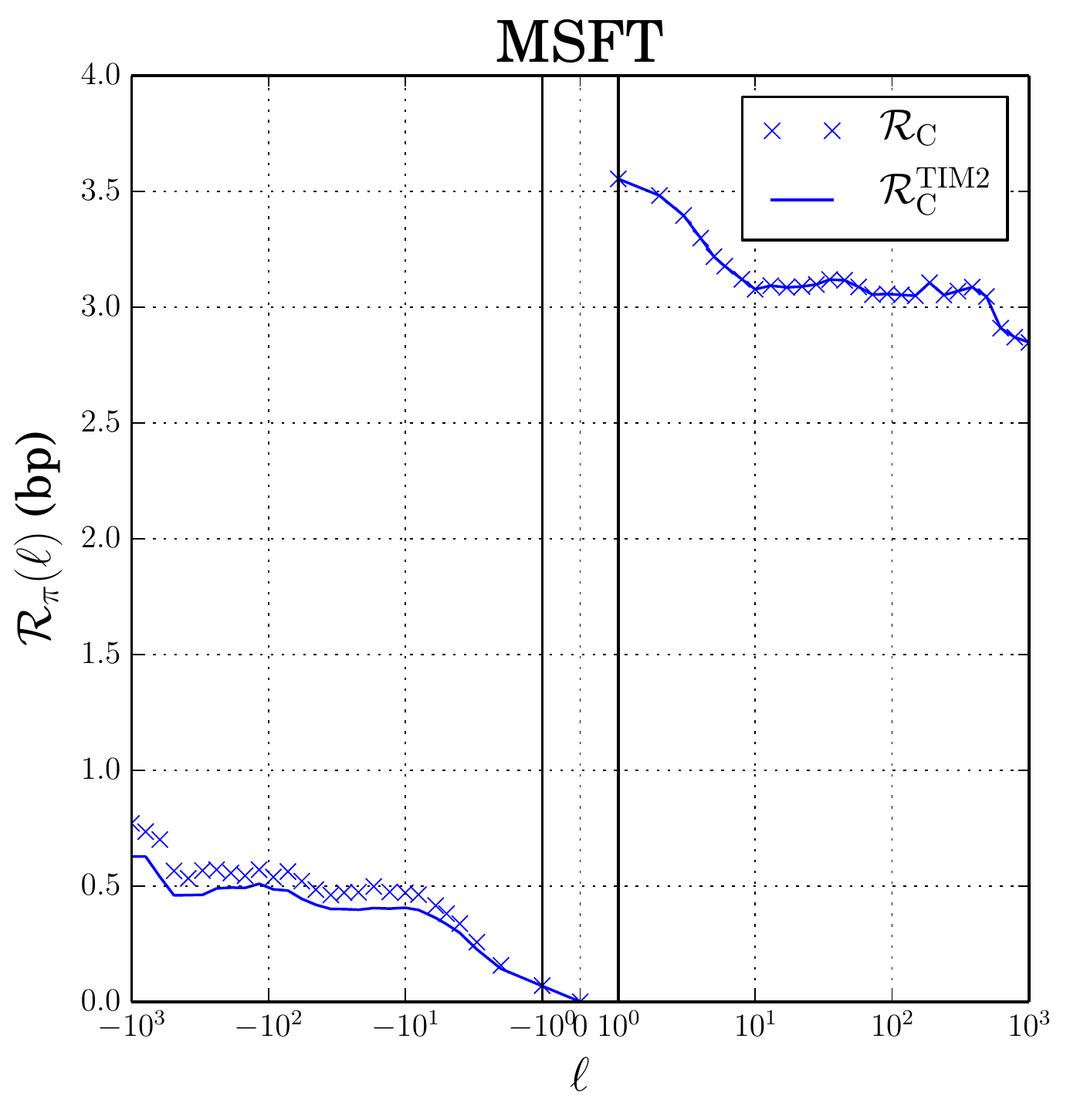}
    \vspace{5mm}
  \end{minipage}%
  \begin{minipage}[c]{0.5\columnwidth}
    \includegraphics[width=\columnwidth]{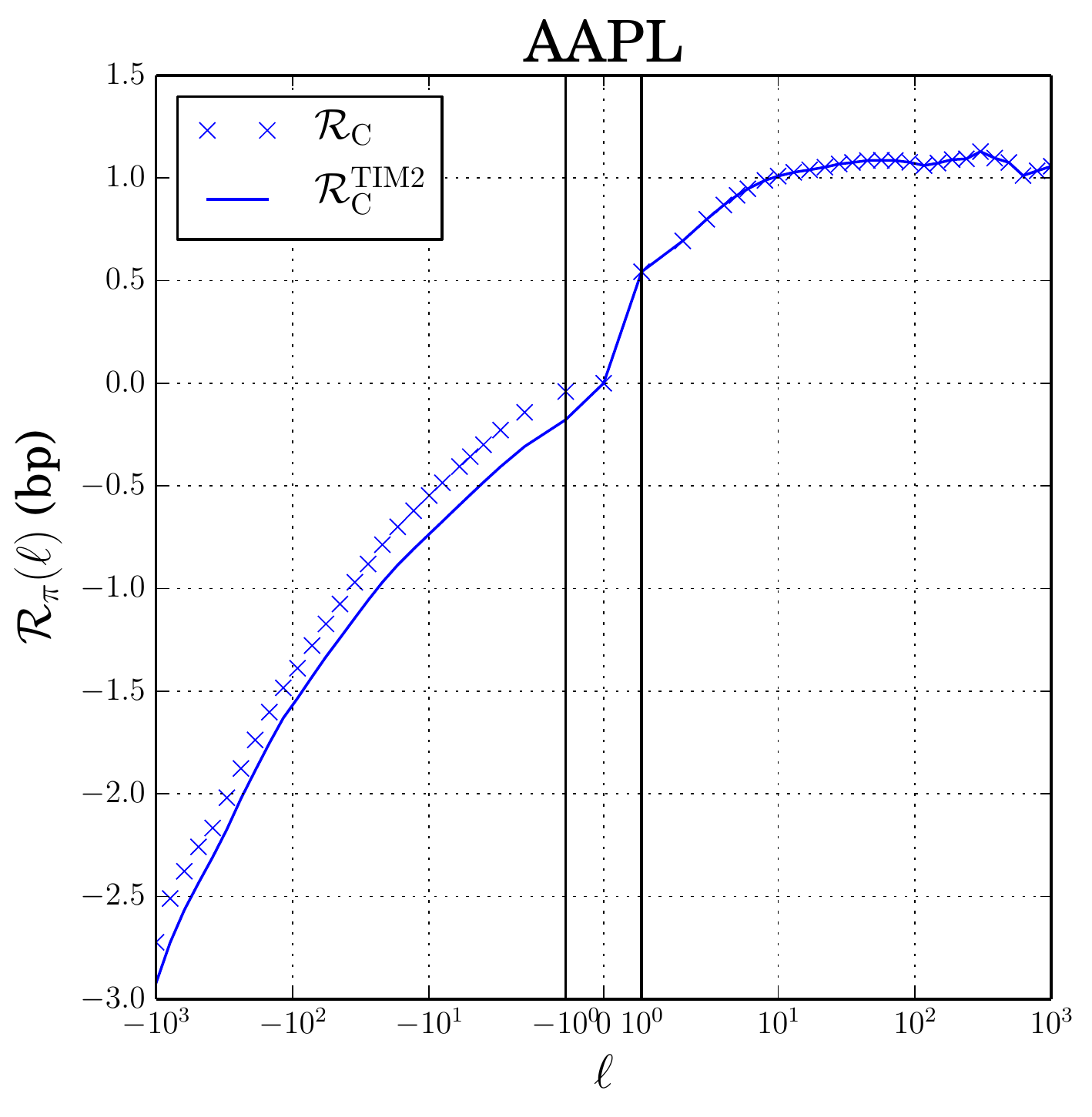}
    \vspace{5mm}
  \end{minipage}
  \caption{(Top panels) Conditional response function for NC events (blue markers) and the theoretical predictions of the TIM2 (solid lines). (Bottom panels) Conditional response function for C events (blue markers) and the theoretical predictions of the TIM2 (solid lines). Left: MSFT (large tick), Right: APPL (small tick). The scale for $\ell$ close to zero and bounded by vertical solid lines is linear, whereas outside this region the scale is logarithmic.}
\label{fig:estimated_resm_rc}
\end{figure}

\begin{figure}[p]
  \begin{minipage}[c]{0.5\columnwidth}
    \includegraphics[width=\columnwidth]{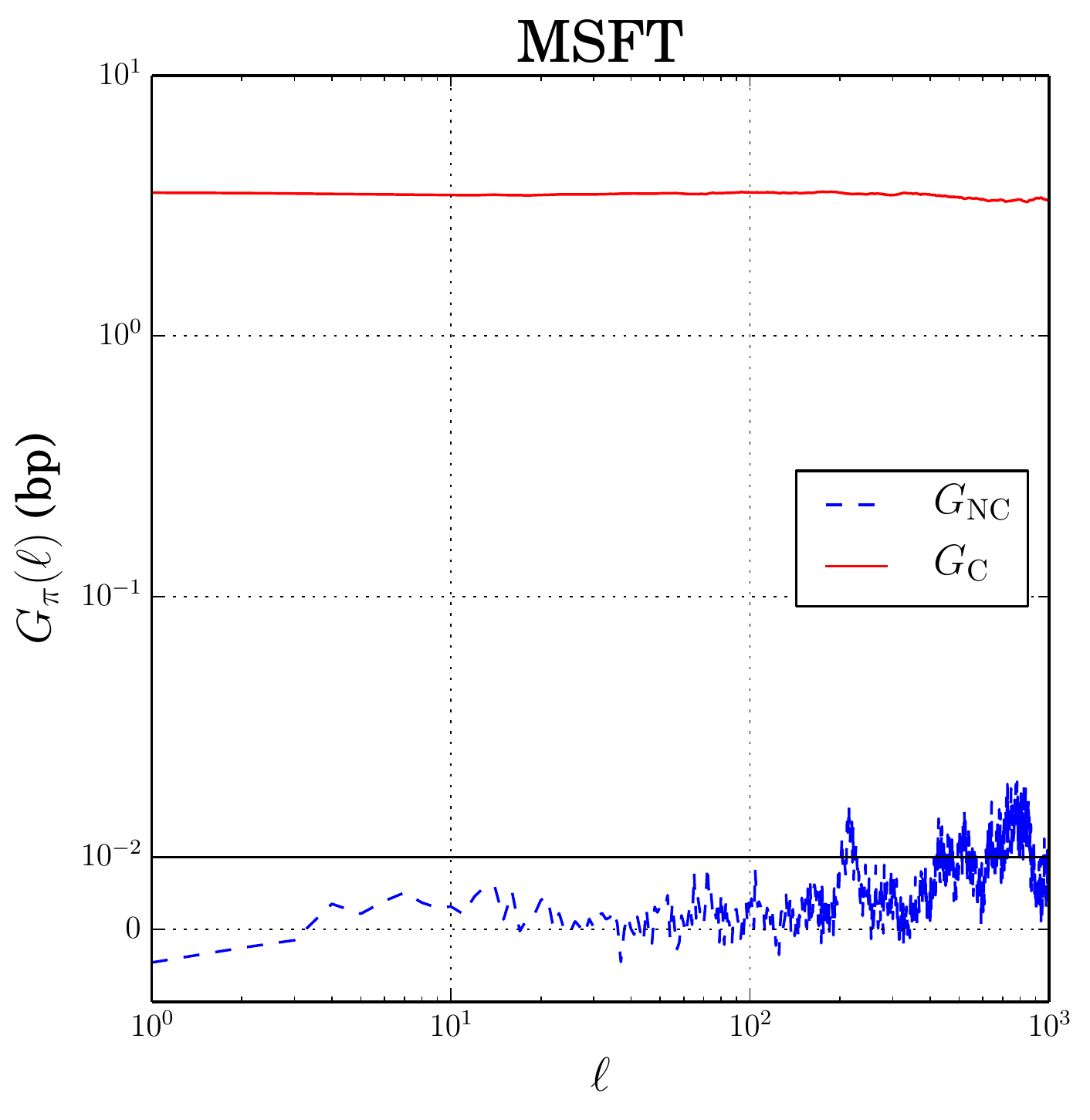}
    \vspace{5mm}
  \end{minipage}%
  \begin{minipage}[c]{0.5\columnwidth}
    \includegraphics[width=\columnwidth]{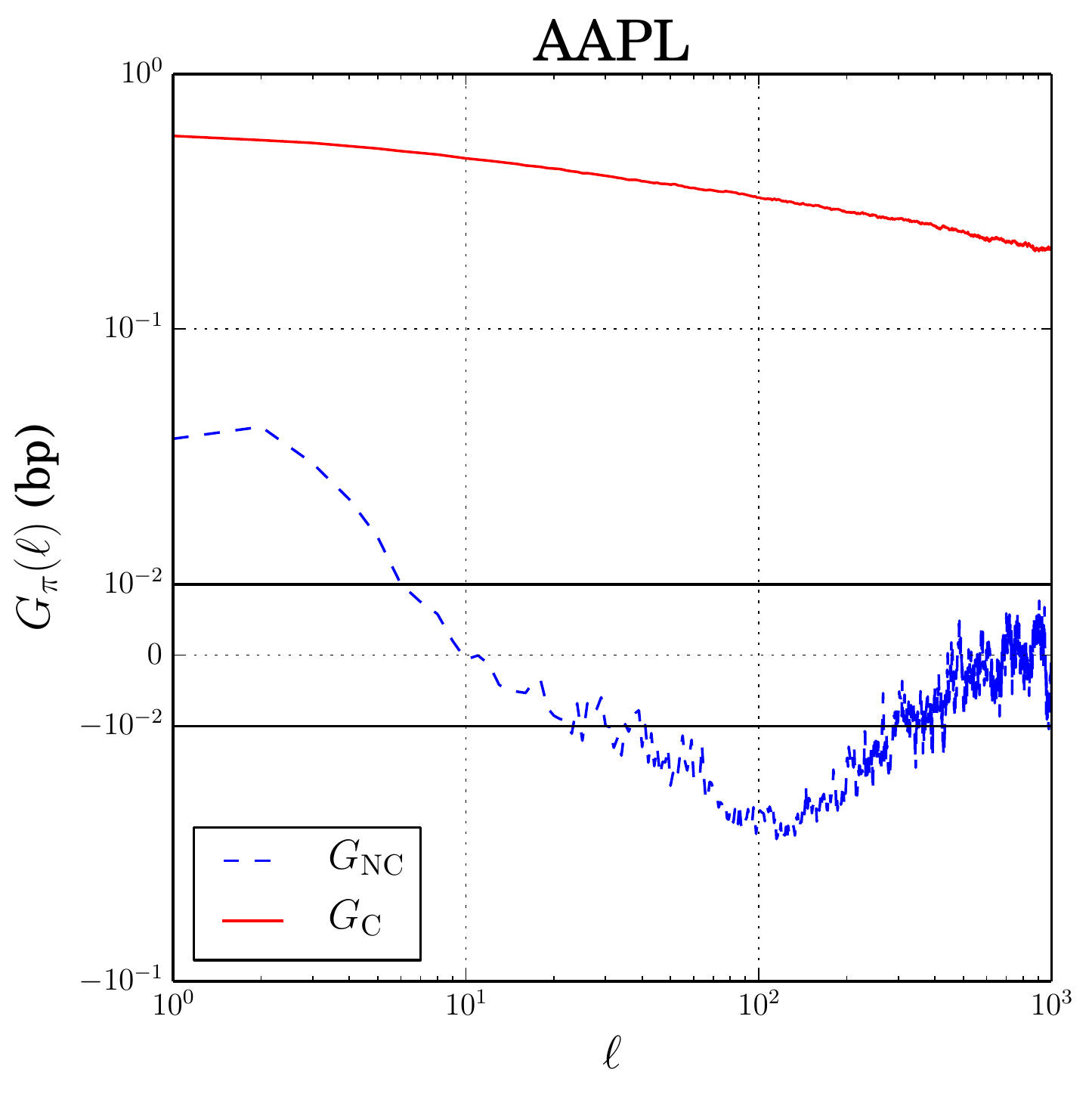}
    \vspace{5mm}
  \end{minipage} \\
  \begin{minipage}[c]{0.5\columnwidth}
    \includegraphics[width=\columnwidth]{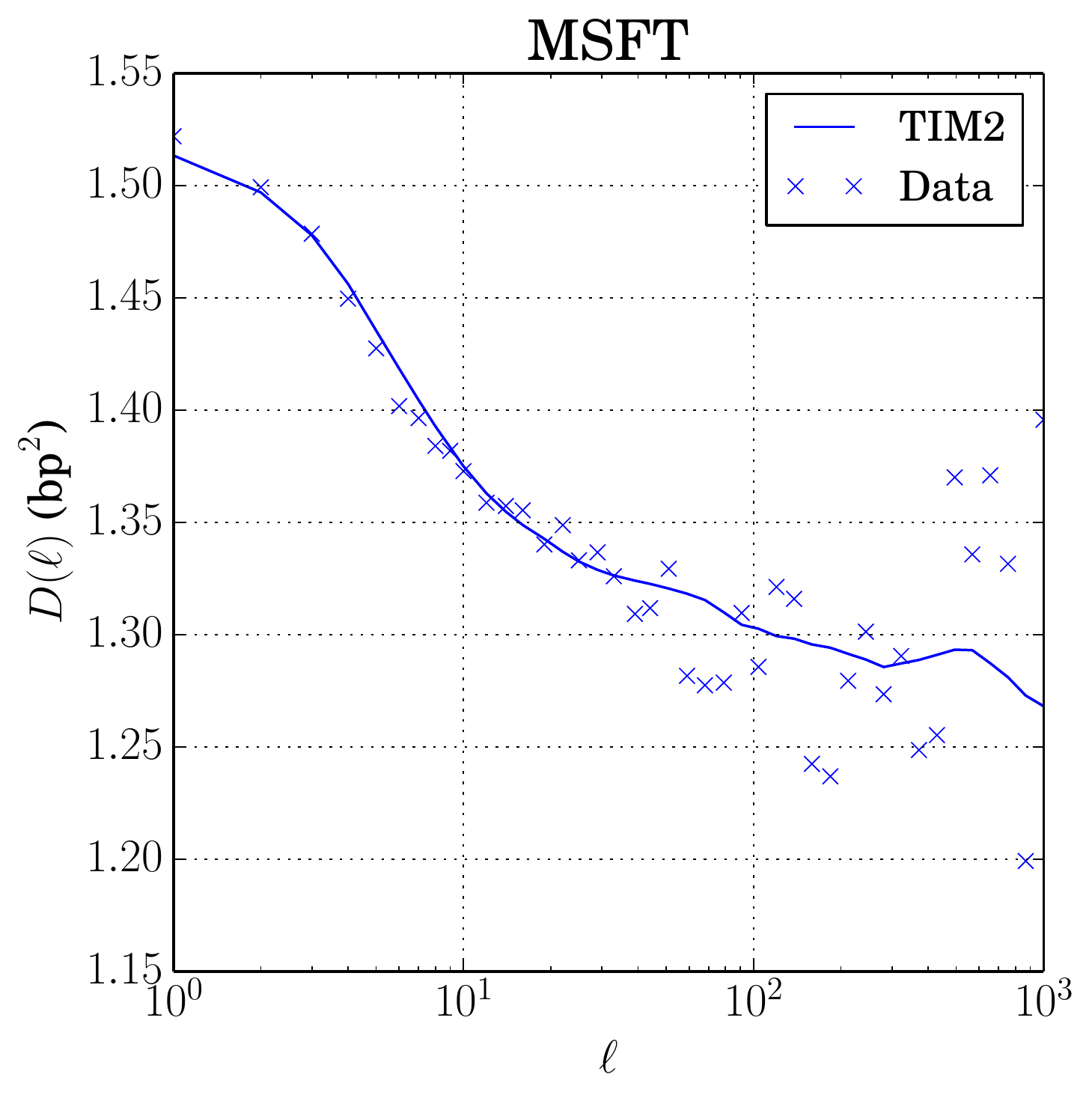}
    \vspace{5mm}
  \end{minipage}%
  \begin{minipage}[c]{0.5\columnwidth}
    \includegraphics[width=\columnwidth]{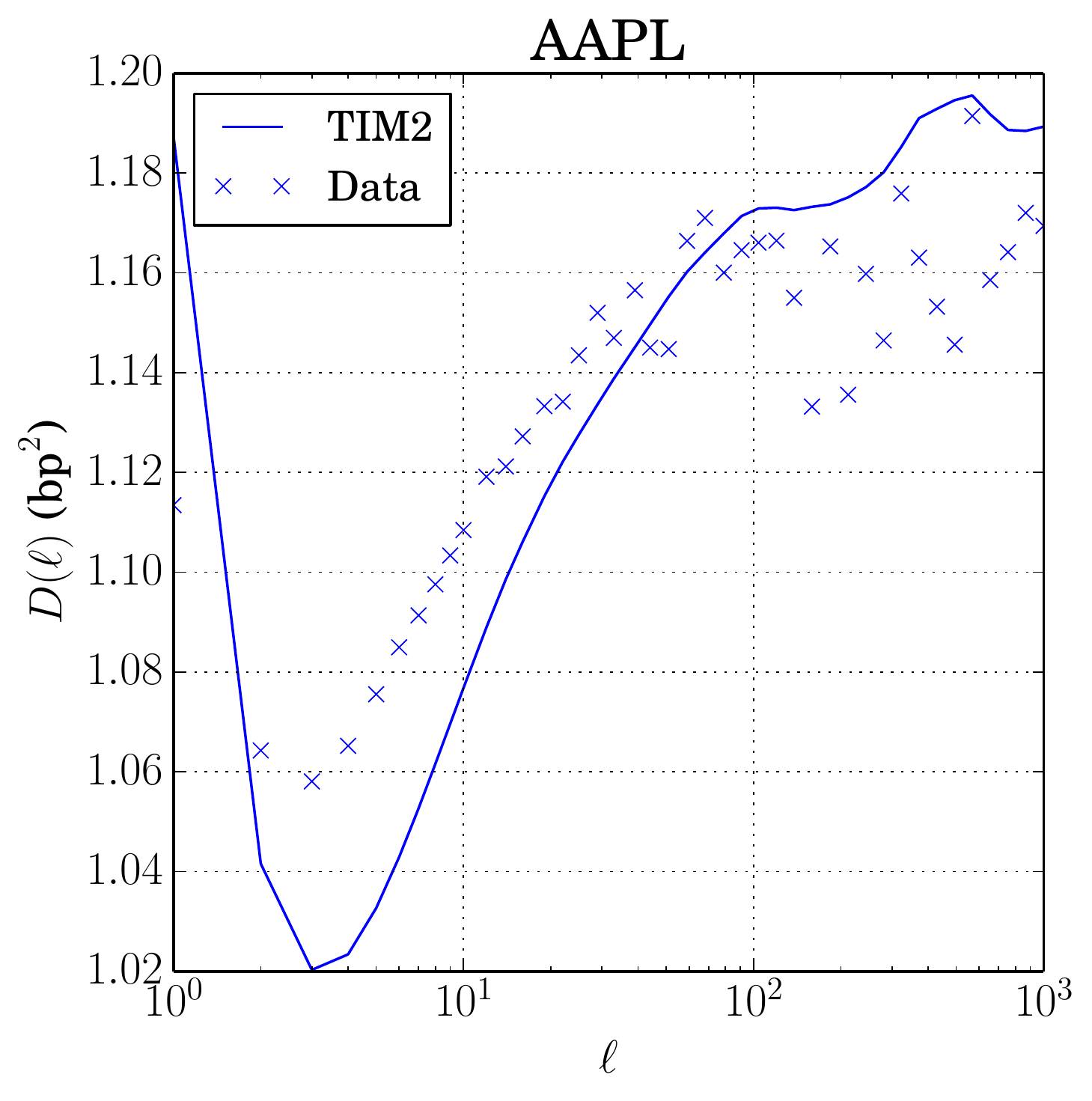}
    \vspace{5mm}
  \end{minipage} 
  \caption{(Top panels) The estimated propagator functions $G_\pi(\ell)$ of TIM2. The scale for $\ell$ close to zero and bounded by horizontal solid lines  is linear, whereas outside this region the scale is logarithmic. (Bottom panels) Signature plots, empirical and predicted by the calibrated TIM2. Left: MSFT with $D_\mathrm{LF}=0.54$ and $D_\mathrm{HF}=0$, Right: APPL with $D_\mathrm{LF}=0.56$ and $D_\mathrm{HF}=0.41$.}
\label{fig:estimated_g_diff_rc}
\end{figure}

Let us now discuss the observed response functions for positive lags, and the resulting calibrated propagators for small tick stocks, as for AAPL, shown in Fig. \ref{fig:estimated_resm_rc} and \ref{fig:estimated_g_diff_rc} (right panels). The conditional response function $\mathcal{R}_\mathrm{C}(\ell)$ after an event of type $\pi=\mathrm{C}$ is a rigid shift of the $\mathcal{R}_\mathrm{NC}(\ell)$ curve. The reaction of market agents to the two types of events is therefore very similar. The shift indeed is due to the very definition of event types, that leads to a non-zero value of $\mathcal{R}_\mathrm{C}(\ell=1)$, comparable to the average spread. Turning now to the conditional response function for negative lags, we observe a small deviation between the model and the empirical data: There exists an additional anti-correlation between past returns and future order signs which is not captured by the model. The curves $\mathcal{R}_\mathrm{NC}(-\ell)$ and $\mathcal{R}_\mathrm{C}(-\ell)$ behave in similar way, but in the latter case the anti-correlation is stronger than in the former case.
The propagator functions $G_\mathrm{C}(\ell)$ can be fit by a power-law, but the $G_\mathrm{NC}$ curves are non monotonic (Fig. \ref{fig:estimated_g_diff_rc}). Note that, as a result of the non-trivial structure of the correlation, the calibration of the TIM2 leads to $G_\mathrm{NC}(\ell=1) > 0$. This is inconsistent with the interpretation of the model -- which would require $G_\mathrm{NC}(\ell=1) = 0$ -- and shows the theoretical limitations of the TIM framework. In the case of the HDIM framework, by construction, we have that $\kappa_{\pi\mathrm{NC}}(\ell=1)=0$.

The results of the estimation of the model for large tick stocks are completely different. Fig. \ref{fig:estimated_resm_rc} and \ref{fig:estimated_g_diff_rc} (left panels) show the results for MSFT. The $\mathcal{R}_\mathrm{NC}(\ell)$ curve is a positive and increasing function which starts, as expected, from zero and reaches a plateau for large lags. The $\mathcal{R}_\mathrm{C}(\ell)$ curve starts from the value of the spread in basis point and slightly decreases, which means that the reaction of the market after price change events consists in a mean reversion of the price. For negative lags, the curve $\mathcal{R}_\mathrm{NC}(-\ell)$ shows that if an event occurs that does not change the price, then for small lags the past returns are on average anti-correlated with the present order sign. The case of the $\mathcal{R}_\mathrm{C}(-\ell)$ is quite interesting, because it shows that if a price changing event occurs, then the past returns are on average anti-correlated with the present order sign.

The propagator functions $G_\pi$ are almost constant with different values: $G_\mathrm{C}$ is equal to the spread, whereas $G_\mathrm{NC}$ is equal to zero. The fact that the two propagators are constant means that the price process in Eq. \ref{eqn:two_prom_price_process} is simply a sum of non-zero price changes, all equal to the spread, and for which the impact is permanent. Therefore, as noted in \cite{eisler2012a} the dynamics of the price is completely determined by the sequence of random variables $\left\lbrace (\epsilon_t,\pi_t) \right\rbrace_{t\in \mathbb{N}}$, and the temporal structure of their correlations. More precisely, if spread fluctuations can be neglected, TIM2 lead to the following simple predictions:
\begin{align}
\mathcal{R}^{\text{TIM2}}_\pi(\ell>0) &\approx \sum_{0 \leq n < \ell} \sum_{\pi_1}\mathbb{P}(\pi_1) G_{\pi_1}(1) C_{\pi, \pi_1}(n) \nonumber \\
&=G_\mathrm{C}(1)\left[1+\sum_{0 < n < \ell} \mathbb{P}(\mathrm{C}) C_{\pi, \mathrm{C}}(n)\right], \nonumber \\
\mathcal{R}^{\text{TIM2}}_{\pi}(\ell<0) &\approx -\sum_{0 < n \leq \ell} \sum_{\pi_1} \mathbb{P}(\pi_1) G_{\pi_1}(1) C_{\pi_1, \pi}(n) \nonumber \\
&=- G_\mathrm{C}(1)\sum_{0 < n \leq \ell} \mathbb{P}(\mathrm{C}) C_{\mathrm{C}, \pi}(n).
\label{eq:rmean}
\end{align}
and:
\begin{align}
D^{\text{TIM2}}(\ell) &\approx D_\mathrm{LF} + \sum_\pi G_\pi(1)^2 \mathbb{P}(\pi) + \frac{2}{\ell} \sum_{0 \leq n < m <\ell} \sum_{\pi_1,\pi_2} \mathbb{P}(\pi_1) \mathbb{P}(\pi_2) G_{\pi_1}(1) G_{\pi_2}(1) C_{\pi_1, \pi_2}(m-n) \nonumber \\
&= D_\mathrm{LF} + G_\mathrm{C}(1)^2 \mathbb{P}(\mathrm{C})+2 \frac{G_\mathrm{C}(1)^2}{\ell} \sum_{0 \leq n < m <\ell}\mathbb{P}(\mathrm{C})^2 C_{\mathrm{C}, \mathrm{C}}(m-n).
\label{eq:vol}
\end{align}
Note that the both the empirical response for negative lags and the signature plot are now perfectly reproduced. The improvement from the TIM1 is quite remarkable.
%**Shouldn't we systematically compare the improvement by plotting a new z* as a function of tick size ? Or would Damian need to come back for this ??** \textit{Yes, it is. I need the data I filtered in Paris last year.}

\subsubsection{Tests on the HDIM2}\label{sec:hdim_estimation} 
The calibration of the HDIM2 model requires the determination of the influence matrix $\kappa_{\pi_1,\pi_2}$, which can be done from the empirical knowledge of the response matrices $\mathcal{S}_{\pi_1,\pi_2}(\ell)$ since
\begin{equation*}
  \mathcal{S}_{\pi_1,\pi_2}(\ell)=G_{\pi_2}(1) C_{\pi_1,\pi_2}(\ell)+\sum_{n>0}\sum_{\pi} \mathbb{P}(\pi)\kappa_{\pi,\pi_2}(n)C_{\pi,\pi_1,\pi_2}(n,\ell),
\end{equation*}
where
\begin{align*}
  \mathcal{S}_{\pi_1,\pi_2}(\ell)&=\frac{\mathbb{E}[I(\pi_{t-\ell}=\pi_1)\epsilon_{t-\ell} \cdot I(\pi_t=\pi_2) r_t]}{\mathbb{P}(\pi_1)\mathbb{P}(\pi_2)}, \\
  C_{\pi,\pi_1,\pi_2}(k,\ell)&=\frac{\mathbb{E}[I(\pi_{t-k}=\pi)\epsilon_{t-k} \cdot I(\pi_{t-\ell}=\pi_1)\epsilon_{t-\ell} \cdot I(\pi_t=\pi_2)]}{\mathbb{P}(\pi)\mathbb{P}(\pi_1)\mathbb{P}(\pi_2)}.
\end{align*}

Actually the previous equation is not convenient to be used for the estimation of the model, because it includes the empirical determination of the three-point correlation functions $C_{\pi,\pi_1,\pi_2}(k,\ell)$. Therefore, in~\cite{eisler2012b} authors employed a Gaussian assumption which leads to the factorization of the three-point correlation functions in terms of two-point correlation functions:
\begin{equation*}
  \mathcal{S}_{\pi_1,\pi_2}(\ell)\approx G_{\pi_2}(1) C_{\pi_1,\pi_2}(\ell)+\sum_{n>0}\sum_{\pi} \mathbb{P}(\pi)\kappa_{\pi,\pi_2}(n)C_{\pi,\pi_1}(n-\ell).
\end{equation*}
The resulting formula for the signature plot $D^{\text{HDIM2}}(\ell)$ is considerably more complicated. We report it for completeness in Appendix~\ref{app:hdim_calibration}. 

\begin{figure}[p]
  \begin{minipage}[c]{0.5\columnwidth}
    \includegraphics[width=\columnwidth]{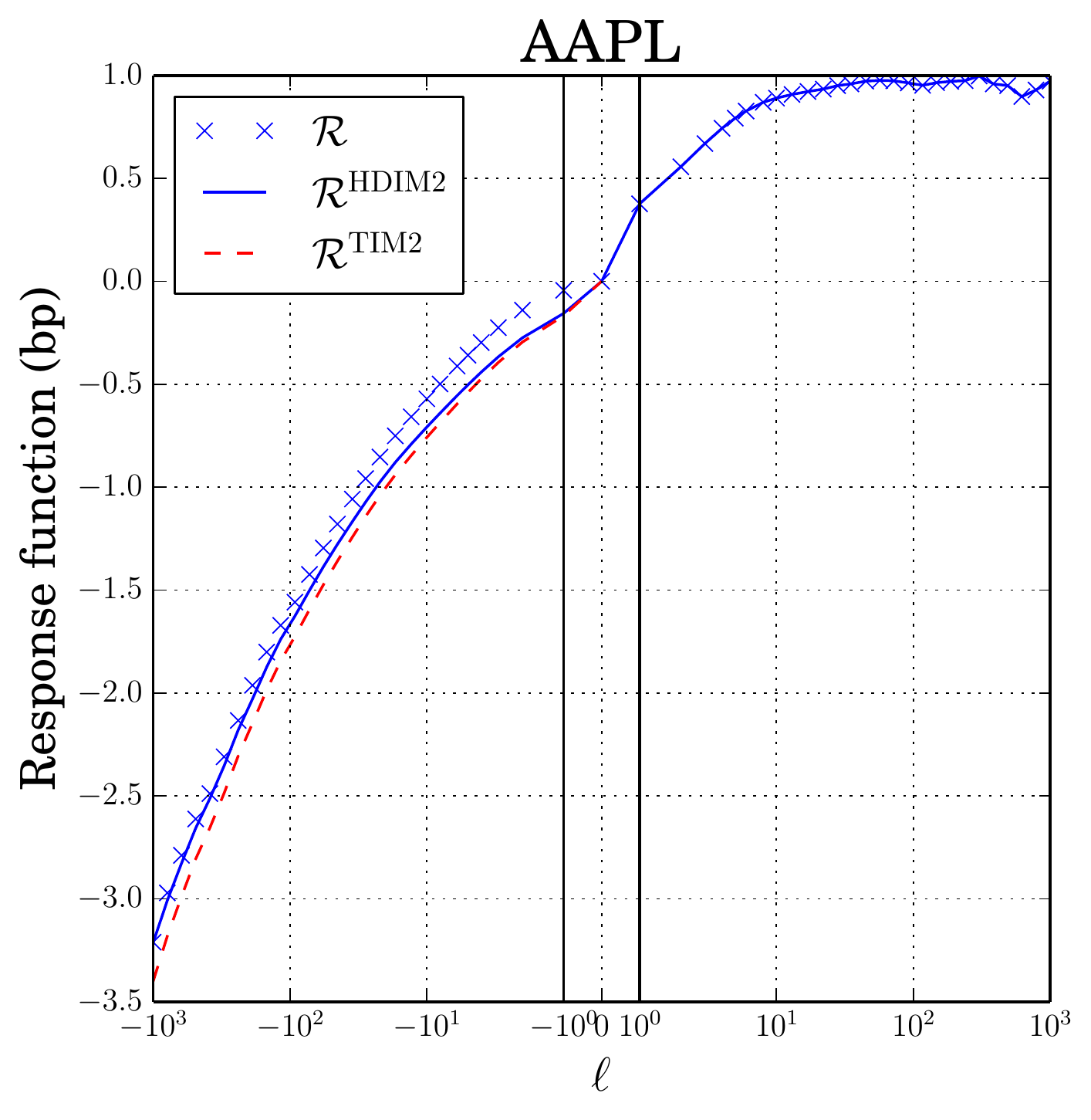}
    \vspace{5mm}
  \end{minipage}%
  \begin{minipage}[c]{0.5\columnwidth}
    \includegraphics[width=\columnwidth]{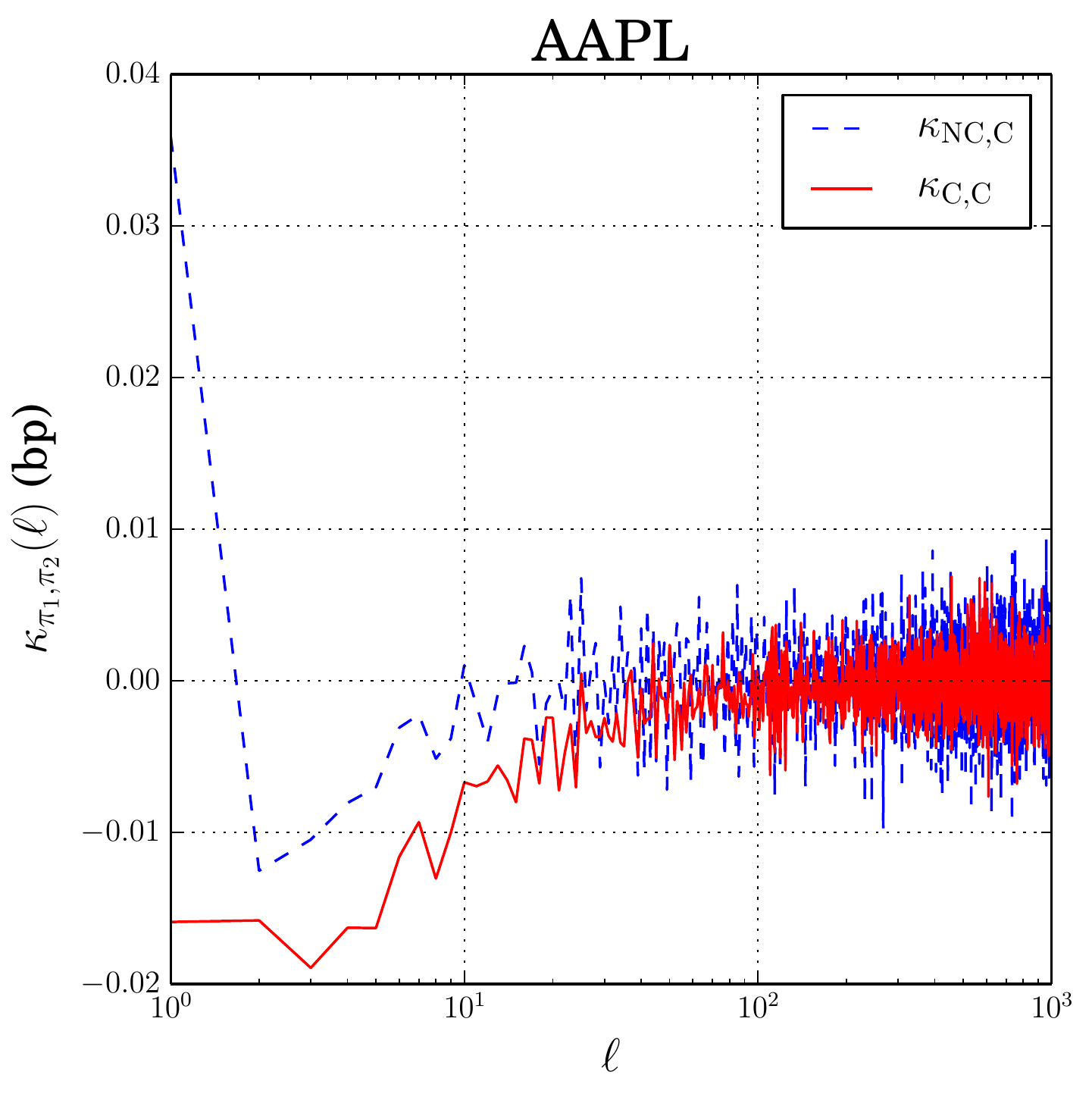}
    \vspace{5mm}
  \end{minipage} \\
  \begin{minipage}[c]{0.5\columnwidth}
    \includegraphics[width=\columnwidth]{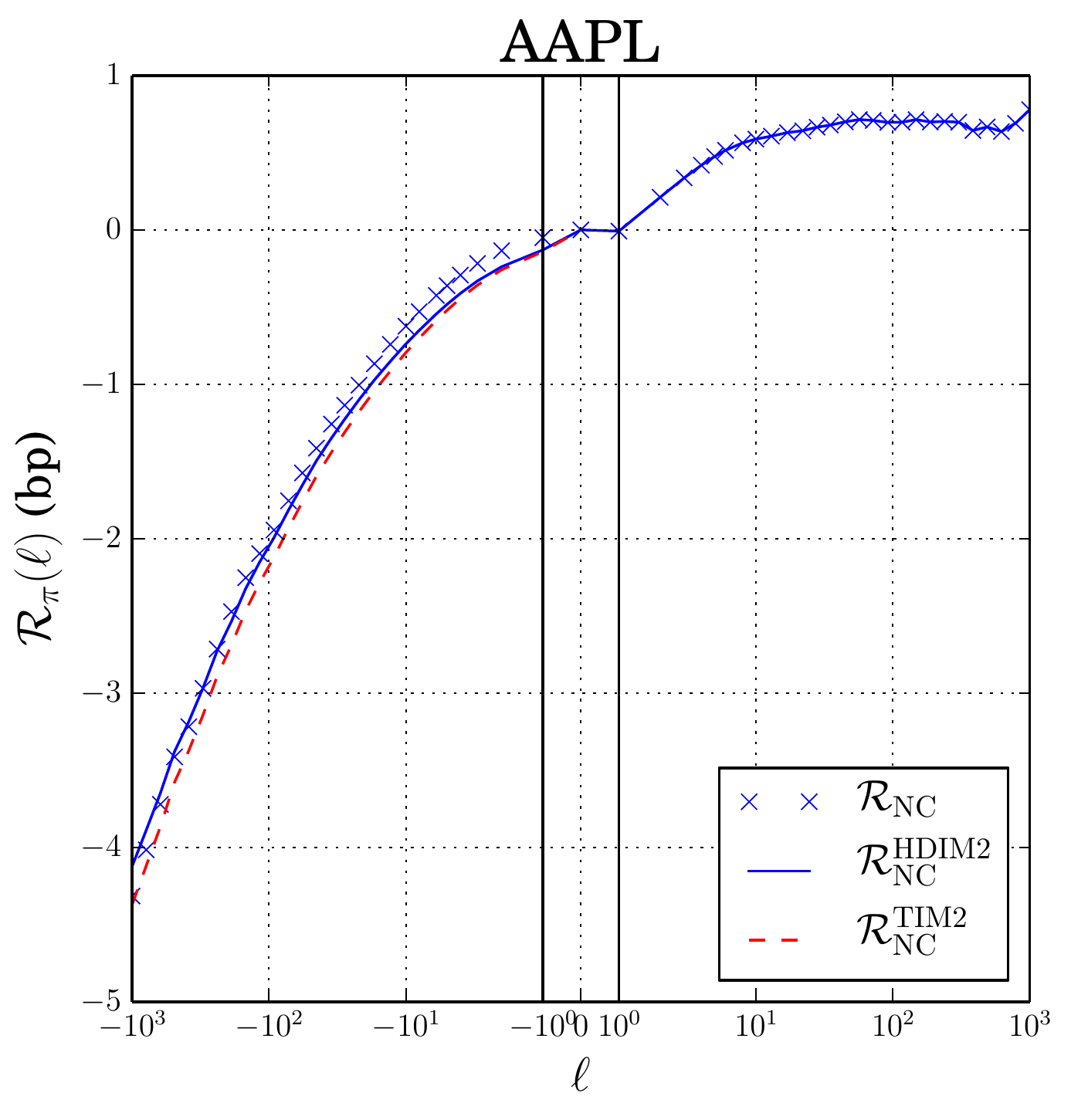}
    \vspace{5mm}
  \end{minipage}%
  \begin{minipage}[c]{0.5\columnwidth}
    \includegraphics[width=\columnwidth]{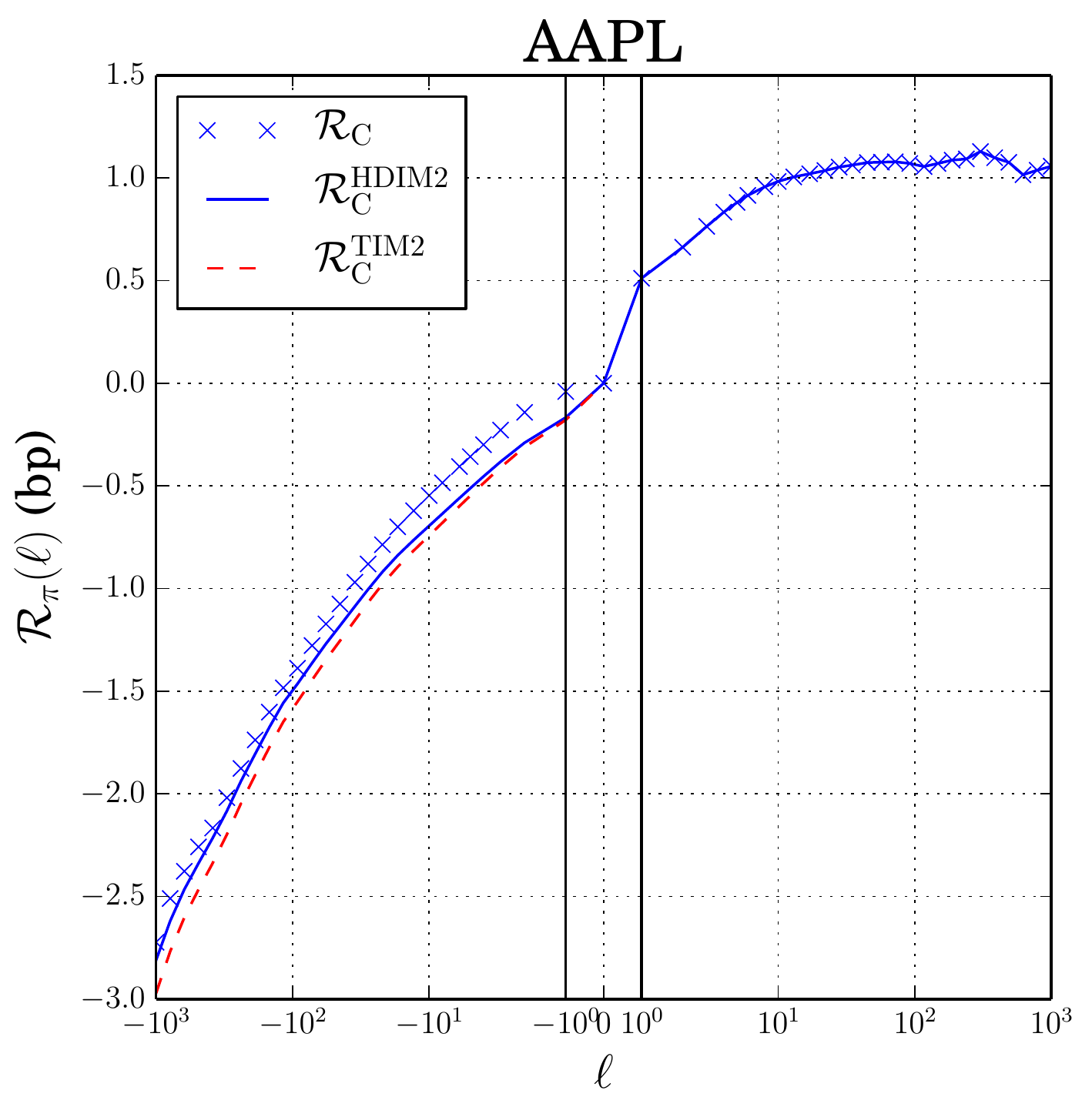}
    \vspace{5mm}
  \end{minipage} 
  \caption{(Top left) Response function (blue markers), theoretical prediction of the HDIM2 (blue solid line), and of the TIM2 (red dashed lines) for AAPL. Top right. Influence kernels $\kappa_{\pi_1,\pi_2}$ of the HDIM2 calibrated on AAPL. (Bottom panels) Conditional response functions (blue markers), theoretical predictions of the HDIM2 (blue solid line), and of the TIM2 (red dashed lines) calibrated on AAPL data. The scale for $\ell$ close to zero and bounded by vertical solid lines is linear, whereas outside this region the scale is logarithmic.}
\label{fig:estimated_g_diff_hdim}
\end{figure}

On purely theoretical grounds, HDIMs are better founded than TIMs and we have extended the above analysis to HDIMs as well. In the case of large tick stocks, there is no gain over the TIM framework since the influence kernels are found to be extremely small. Any gain is therefore only possible for small tick stocks. We show the empirical determination of the two influence kernels $\kappa_{\pi_1,\mathrm{C}}(\ell)$ as well as the resulting predicted response ${\cal R}^{\text{HDIM2}}_{\pi}(\ell)$ for AAPL in Fig. \ref{fig:estimated_g_diff_hdim}. As can be noted, the estimated kernels differ whether the sequence of events which precede the price-changing trade is composed of price-changing or non price-changing orders. We can argue that Eq.~\ref{eqn:hdim_tim_model} -- which neglects the role of the realised event -- is too restrictive. It is worth to comment that, when statistically different from zero, the influence kernel $\kappa_{C,C}$ is negative. Then, a sequence of price-changing orders on the same side of the final C trade is going to impact the market less than a C order preceded by a sequence of price-changing events of the opposite sign. Thus we see the same asymmetric liquidity mechanism described in~\cite{lillo2004}. As a sole difference with the picture described in section~\ref{sec:propagator}, the influence kernel  $\kappa_{NC,C}$ is positive for the very last NC event occurring before a price-changing event. This implies that the impact of the C market order is larger if it follows a sequence of NC trades whose last event occurs on the same side of the C event. 

We see some further improvement over the TIM2 for the conditional response functions at negative lags. It seems that HDIM2 performs slightly better than TIM2 in capturing the excess anti-correlation measured from the data between past returns and future order signs. We also observe an improvement -- albeit in a marginal way -- for the signature plot in Fig. \ref{fig:estimated_g_diff_hdim}. We recall here that in the 6-event extension of the propagator model considered in \cite{eisler2012b}, HDIMs appeared to fare slightly worse than TIMs for small tick stocks, for a reason that is still not well understood, and that would 
deserve further scrutiny.

\begin{figure}[t]
\centering
\includegraphics[width=0.5\columnwidth]{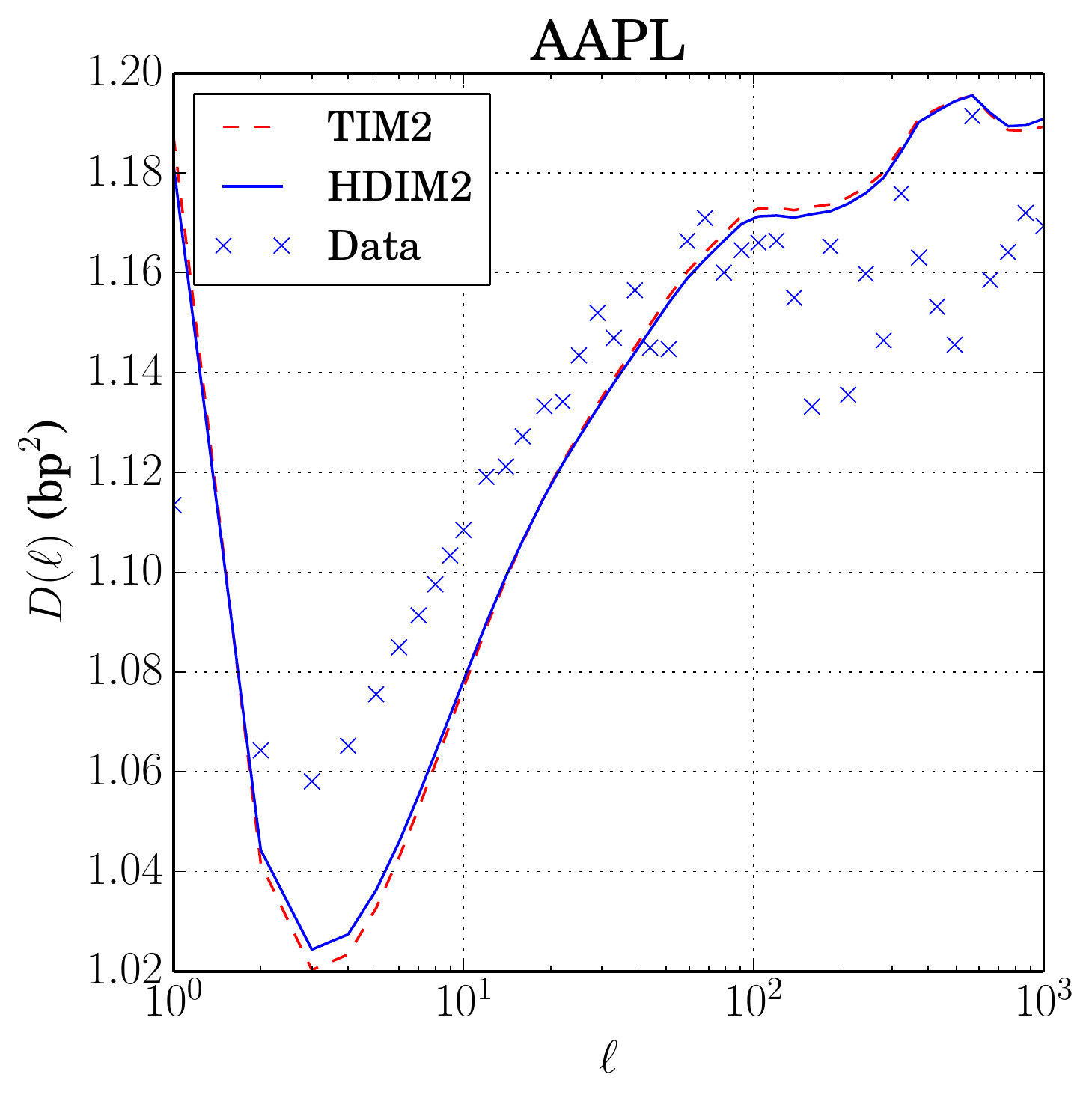}
\caption{Signature plots for AAPL, namely empirical, predicted by the HDIM2 (with $D_\mathrm{LF}=0.6$ and $D_\mathrm{HF}=0.39$), and by the TIM2.}
\label{fig:signature_plot_hdim}
\end{figure}

\section{Discussion and (partial) conclusion}
The above study attempts to build the most accurate linear model of price dynamics based on the only observation of market orders. 
We have seen that treating all market orders on the same footing, as in the first version of the propagator model, leads to systematic discrepancies that increase with the tick size. For large tick sizes, the predictions of this simple framework are qualitatively erroneous, both for the price response at negative lags and for the diffusion properties of the price. This can be traced to the inability of the model to describe the feedback of price changes on the order flow, which is strong for large tick stocks. Generalizing the model to two types of market orders, those which leave the price unchanged and those which lead to an immediate price change, considerably improves the predictive power of the model, in particular for large ticks for which the above inadequacy almost entirely disappears, leading to a remarkable agreement between the model's predictions and empirical data. We have also seen that, although better justified theoretically, the ``history dependent'' impact models (HDIM) fare only slightly better than the ``transient'' impact models (TIM) when only two event types are considered.

Still, we are left with two important questions about the order flow itself, which we considered ``rigid'' in the above formalism, in the sense that it is entirely described by its correlation structure and does not explicitly react to past events (at variance with the price itself). It would be desirable to develop a more dynamic description of the order flow, for at least two reasons. One is that linear models are best justified in a context where the best predictor of the order flow is itself linear, as is the case of DAR processes for the sign of market orders. We therefore need to generalize DAR processes to a multi-event context, and see how well the corresponding so-called MTD models account for the statistics of the order flow, i.e. the string of $\lbrace(-1,\mathrm{C})\rbrace$, $\lbrace(-1,\mathrm{NC})\rbrace$, $\lbrace(+1,\mathrm{NC})\rbrace$, $\lbrace(+1,\mathrm{C})\rbrace$ events. The second reason is that the ``true'' impact of an additional market order, not present in the past time series, should include the mechanical contributions captured by the TIMs or HDIMs, but also the possible change of the order flow itself due to an extra order in the market, an effect clearly not captured by our assumption of a rigid order flow. We thus need to define and calibrate the equivalent of the influence kernels defined above, but for the order flow itself. This is what we  do in the following companion paper.

\section*{Acknowledgement}
We thank I. Mastromatteo, J. Donier, J. Kockelkoren and especially Z. Eisler for many inspiring discussions on these topics.

\appendix
\section{Diffusion properties of TIMs}
\label{app:tim_signature_plot}
The exact expression of the diffusive curve $D(\ell)$, given in \cite{eisler2012a}, is:
\begin{align*}
  D^\mathrm{TIM}(\ell)\ell&=D_\mathrm{LF}\ell+D_\mathrm{HF}+\sum_{0 \leq n < \ell}\sum_\pi \mathbb{P}(\pi) G_\pi(\ell-n)^2+\sum_{n>0}\sum_\pi\mathbb{P}(\pi)\left[G_\pi(\ell+n)-G_\pi(n)\right]^2 \\
  &+2\sum_{0 \leq n < m < \ell} \sum_{\pi_1,\pi_2} \mathbb{P}(\pi_{1,2}) G_{\pi_1}(\ell-n)G_{\pi_2}(\ell-m) C_{\pi_1,\pi_2}(m-n) \\
  &+2\sum_{0 \leq n < m} \sum_{\pi_1,\pi_2} \mathbb{P}(\pi_{1,2}) \left[G_{\pi_1}(\ell+n)-G_{\pi_1}(n)\right]\left[G_{\pi_2}(\ell+m)-G_{\pi_2}(m)\right]C_{\pi_2,\pi_1}(m-n) \\
  &+\sum_{0 \leq n < \ell} \sum_{m>0} \sum_{\pi_1,\pi_2} \mathbb{P}(\pi_{1,2}) G_{\pi_1}(\ell-n)\left[G_{\pi_2}(\ell+m)-G_{\pi_2}(m)\right]C_{\pi_2,\pi_1}(m+n),
\end{align*}
where $\mathbb{P}(\pi_{i,\ldots,j})=\mathbb{P}(\pi_i)\cdots\mathbb{P}(\pi_j)$.

\section{Diffusion properties of HDIMs}
\label{app:hdim_calibration}
Knowing the $\kappa_{\pi_1,\pi_2}$'s and using the factorization of three-point and four-point correlations in terms of two-point correlations, one can finally estimate the diffusion curve, which is given by the following approximate equation:
\begin{align*}
  D^\mathrm{HDIM}(\ell)\ell&\approx\bigg[D_\mathrm{LF}+\frac{D_\mathrm{HF}}{\ell}+\sum_{\pi_1}G_{\pi_1}(1)^2\mathbb{P}(\pi_1)+\sum_{n>0}\sum_{\pi_1,\pi_2}\mathbb{P}(\pi_{1,2})\kappa_{\pi_1,\pi_2}(n)^2[\Pi_{\pi_1,\pi_2}(n)+1] \\
  &+2\sum_{0<n<m}\sum_{\pi_1,\pi_2,\pi_3}\mathbb{P}(\pi_{1,2,3})\kappa_{\pi_1,\pi_3}(m)\kappa_{\pi_2,\pi_3}(n)C_{\pi_1,\pi_2}(m-n) \\
  &+2\sum_{n>0}\sum_{\pi_1,\pi_2} G_{\pi_2}(1) \kappa_{\pi_1,\pi_2}(n) C_{\pi_1,\pi_2}(n) \bigg]\ell \\ 
  &+2\sum_{0 < n < \ell}\sum_{\pi_1,\pi_2}(\ell-n)\mathbb{P}(\pi_{1,2})G_{\pi_1}(1)G_{\pi_2}(1) C_{\pi_1,\pi_2}(n) \\
  &+2\sum_{0 < n < \ell} \sum_{i>0} \sum_{\pi_1,\pi_2,\pi_3}(\ell-n) \mathbb{P}(\pi_{1,2,3}) G_{\pi_1}(1) \kappa_{\pi_2,\pi_3}(i) C_{\pi_1,\pi_2}(n+i) \\
  &+2\sum_{0 < n < \ell} \sum_{\substack{i>0 \\ i \neq n}} \sum_{\pi_1,\pi_2,\pi_3}(\ell-n) \mathbb{P}(\pi_{1,2,3}) G_{\pi_1}(1) \kappa_{\pi_2,\pi_3}(i) C_{\pi_1,\pi_2}(n-i) \\
  \qquad &+2\sum_{0 < n < \ell}\sum_{\pi_1,\pi_2}(\ell-n) \mathbb{P}(\pi_{1,2}) G_{\pi_1}(1) \kappa_{\pi_1,\pi_2}(n) [\Pi_{\pi_1,\pi_2}(n)+1] \\
  &+2\sum_{0 < n < \ell} \sum_{\substack{i,j>0 \\ j \neq n}} \sum_{\substack{\pi_1,\pi_2 \\ \pi_3,\pi_4}} (\ell-n) \mathbb{P}(\pi_{1,2,3,4}) \kappa_{\pi_1,\pi_2}(i) \kappa_{\pi_3,\pi_4}(j) C_{\pi_1,\pi_3}(n+i-j)[\Pi_{\pi_2,\pi_4}(n)+1] \\
  &+2\sum_{0 < n < \ell} \sum_{i>0} \sum_{\pi_1,\pi_2,\pi_3}(\ell-n) \mathbb{P}(\pi_{1,2,3}) \kappa_{\pi_1,\pi_2}(i) \kappa_{\pi_2,\pi_3}(n) C_{\pi_1,\pi_2}(i),
\end{align*}
where $\mathbb{P}(\pi_{i,\ldots,j})=\mathbb{P}(\pi_i)\cdots\mathbb{P}(\pi_j)$ and
\begin{equation*}
  \Pi_{\pi_1,\pi_2}(\ell)=\frac{\mathbb{E}[I(\pi_{t-\ell}=\pi_1) \cdot I(\pi_t=\pi_2)]}{\mathbb{P}(\pi_1)\mathbb{P}(\pi_2)}-1.
\end{equation*}

\end{document}